\documentclass[twocolumn]{aastex631}

\usepackage{url}

\usepackage{amsmath}
\usepackage{float}
\usepackage{graphicx}
\usepackage{multirow}
\usepackage{soul}
\usepackage{natbib}
\usepackage{listings}
\usepackage{framed}
\usepackage{pgfplotstable}
\usepackage{booktabs}

\begin{document}
\title{Highly Efficient Identification of Extreme Emission Line
Galaxies in the Local Universe:\\ $>$8000 New Green Pea Candidates at $0.12 < z < 0.36$}
\author{Heather Samonski}
\author{Samir Salim}
\author{John Salzer}
\affil{Department of Astronomy, Indiana University, Bloomington, IN 47408, USA}

\begin{abstract}
The currently known compact extreme emission-line galaxies (the ``Green Peas", GPs) in SDSS are rare and were mostly found among serendipitous spectroscopic targets, thus leaving open the possibility that a substantial population of GPs is missed. A significantly larger number of identified GPs in the Local Universe might provide a better characterization of their high-redshift analogs and Lyman continuum escape. In this paper, we confront the challenges of robustly identifying GPs without spectroscopic information, a needed approach considering the incompleteness of spectroscopic surveys for compact sources. The principal difficulty stems from a significant contamination of photometric candidates by stars and quasars of similar color. To solve this, we introduce an SED matching method, which separates candidate GPs from contaminants on the basis of SDSS and WISE photometry of spectroscopically confirmed stars, quasars and galaxies. The method has an effectiveness of 85\%, and a contamination rate of $\sim$10\%. With it we identify $\sim$9600 GP candidates expected to lie in the $0.12 < z < 0.36$ range---a tenfold increase over what would be selected using SDSS DR18 spectra.  Some of the new GPs are as bright as $r \sim 19$, and 1200 are predicted to have [OIII]5007 equivalent widths in excess of 500 \AA. The new population contains many ``Extended Peas", which are absent among known GPs and possibly represent merging systems. We provide catalogs containing 8313 newly identified GP candidates, as well as 917 GPs confirmed using SDSS spectroscopy and 521 GPs with spectroscopic redshifts from LAMOST and other sources. 

\end{abstract}

\section{Introduction}
Green Peas are compact, low-redshift starbursting galaxies that were originally discovered by the Galaxy Zoo project \citep{Lintott08} volunteers as unresolved green-looking objects in Sloan Digital Sky Survey (SDSS) images. The Peas appear green in SDSS color composite images (where $gri$ bands are mapped as blue-green-red) due to the very strong [OIII] $\lambda$5007 Å emission line that gets shifted into the SDSS $r$-band filter at $0.1 \lesssim z \lesssim 0.4$. As a result, GPs have much bluer $r-i$ or $r-z$ color than regular galaxies. The Green Peas (GPs) were originally presented and characterized in \cite{Cardamone09} (hereafter C09), who used SDSS spectroscopy coupled with color cuts to select them. C09 found GPs to be rare, low- to moderate-mass galaxies $(10^{8.5-10}M_\odot)$ residing in low-density environments, having high star formation rates for their mass, high [OIII]5007 Equivalent Widths (rest-frame EWs, of up to 1000 Å or more), low metallicities (log[O/H] + 12 $\sim 8.1$; \citealt{Amorin10}), and low internal dust reddening ($E(B - V) \leq 0.25$). 

Owing to these properties and also the large ionization parameter (i.e., [OIII] $\lambda$5007/[OII] $\lambda$3727 ratio), the GPs are thought to be low-redshift analogs of the galaxies that reionized the intergalactic medium (IGM) by leaking ionizing radiation (e.g., \citealt{Izotov18,Yang17,Bassett19,Flurry22}). Thus, GPs offer an indirect probe into the reionization era, marking the formation of the first stars and galaxies. 

Various studies after C09 have identified additional GPs, as well as their counterparts that reside below and above the original redshift selection window \citep{Izotov11, Amorin15, Yang17, LiMalkan18, Jiang19, Brunker20, Liu22, Boyett24}. The Peas are also detected in surveys of extreme emission line galaxies (EELGs; \citealt{Boyett24}, \citealt{Lumbreras-Calle22}, \citealt{Iglesias-Paramo22}, \citealt{Breda22}, \citealt{Perez-Montero21}, \citealt{Brunker20}), compact star-forming galaxies \citep{Izotov16,Ding23},  blue compact dwarf galaxies (BCDs; \citealt{Janowiecki14,Laufman22,Monreal-Ibero23, Henkel22}), extremely metal poor galaxies (XMPs; \citealt{SanchezAlmeida16}, \citealt{Senchyna19}, or strong H$\alpha$ emitters (HAEs; \cite{Shim13}). Despite this seeming diversity, the essential defining observable features of GPs are compactness and high EWs, especially of the [OIII]5007 line. 

C09 identified 251 GPs from an area that they estimated as 8000 sq.\ deg., giving a very low surface density of 0.03 Peas per sq.\ deg. C09 also estimated that the true density of GPs might be 50 or so times higher (2 per sq.\ deg.), but without being able to identify this larger population. Why is there such a huge discrepancy between the two values? The very blue $r-z$ color forms the basis of GP selection, since it separates GPs from much more common regular galaxies (see Fig.\ 2 in C09). However, spectroscopy is needed to confirm that the object is a galaxy rather than a Galactic star or a distant quasar of similar color. C09 used existing SDSS spectroscopy to identify GPs, which mostly came from the \textit{serendipitous} component of SDSS spectroscopic targeting, aiming at point sources with unusual colors only when spare fibers were available. Thus, the serendipitous survey was not meant to be complete. Subsequent efforts, in particular with the LAMOST spectroscopic survey, have more specifically targeted potential GPs, quadrupling the number of confirmed objects lying in the $0.1 \lesssim z \lesssim 0.4$ window \citep{Liu22,Liu23}. However, even with this increase, we are in all likelihood still far from having identified most of the GPs, even if considering only the brighter, more massive ones. The goal of the current study is to remedy this situation and increase, potentially significantly, the number of GPs at low redshifts.

Efforts have been made to identify GPs using surveys besides SDSS and LAMOST, using deeper and more complete spectroscopy \citep{Amorin15}, or using narrow-band imaging \citep{Brunker22,Kimsey-Miller24}. Such studies offer an opportunity to establish the true volume density of GPs. However, since these efforts are limited in area, they cannot provide a fuller census of GPs across the sky. 

Going beyond the current situation where we are missing a huge fraction of GPs introduces challenges. Currently, the identification of GPs relies on spectroscopy, as it is considered necessary to confirm the identity of compact, color-selected objects as galaxies and rule out contamination \citep{Lumbreras-Calle22, Senchyna19}. However, it is very time consuming to target all compact-looking objects with unusual colors, which means that obtaining a relatively complete census of GPs using spectroscopy is unrealistic.  Although some studies have obtained quite pure photometrically-selected samples of the most extreme GP-like objects at very low redshifts ($\mathrm{EW[O\,III]}>800$\AA, $z<0.05$; \citealt{Yang17}), the more complete photometrically-selected samples of a broader population of GPs in the redshift window in which the GPs were originally identified ($0.1 \lesssim z \lesssim 0.4$) remains unavailable. This is what our study aims to address.

To approach the problem of photometrically selecting GPs, we will first analyze spectroscopically selected GPs and other objects of similar color (the contaminants) for which spectra are available, and then perform a new method that we introduce (the SED matching method) to obtain a reasonably robust set of GP candidates. Our new larger sample will help us understand just how rare GPs are, and whether the properties of the larger GP population go beyond those of the currently known GPs. A significantly larger GP population, in turn, will allow for better characterizations of their high-redshift analogs and prepare us for the exploitation of future large photometric surveys, such as Euclid and LSST.

\begin{table*}
    \centering
    \begin{tabular*}{\linewidth}{|p{0.02\linewidth}|p{0.13\linewidth}|p{0.255\linewidth}|p{0.28\linewidth}|p{0.03\linewidth}|p{0.163\linewidth}|}
        \hline
        \textbf{No}	&	\textbf{Sample}	&	\textbf{Description}	&	\textbf{Purpose}	&	\textbf{Sec.}    &   \textbf{Number of objects}\\
        \hline
        1	&	Color-selected sample	&	C09-like color-selected sample of objects with galaxy spectra and redshifts in the $0.1<z<0.4$ window	&	Starting point for spectroscopic selection and for the refinement of photometric selection, using EW information  &   \S\ \ref{sec: initial sample}   &   1022 (of which 253 have [OIII]5007 EW measurements) \\
        \hline    
        2	&	Objects with any spectroscopic class and any redshift	&	Color-selected objects that include GPs as well as their contaminants (stars, QSOs, non-GP galaxies)	&	Construction of additional cuts that photometrically remove contaminants from outside of the GP redshift range	(mostly stars and $z\lesssim0.03$ HII regions) &	\S\ \ref{sec: Prelim Sample}   &	4157 \\
        \hline    
        3	&	Spectroscopically confirmed GPs	&	Subset of Sample 1 that passes additional cuts identified using Samples 1 and 2 that remove non-GP galaxies	&	Final, clean sample of spectroscopically confirmed GPs from SDSS	&	\S\ \ref{sec:cat}   &	917 (Table \ref{tab:DR18 GPs}). We refer to a subset of 214 GPs that would have been available to C09 as DR7 GPs. \\
        \hline    
        4	&	Reference set	&	Subset of Sample 2 that passes additional photometric cuts identified using Samples 1 and 2	&	Objects of known nature (GP, star, QSO) to serve as classification reference for objects of unknown type (the candidate pool, Sample 5)	&	\S\ \ref{sec: Refined sample}   &	3133 \\
        \hline    
       5	&	Candidate pool	&	All objects (in particular those without spectra) that pass all photometric cuts above as well as the positional cuts (to exclude the MW band)	&	Objects that are subjected to the SED matching method in order to separate GP candidates (Sample 6) from likely contaminants (mostly quasars and stars)	&	\S\ \ref{sec:final}   &	18,880 (SQL code given in Appendix \ref{app}) \\
        \hline    
       6	&	GP candidates	&	Subset of candidate pool (Sample 5) classified as GPs by the SED matching method	&	The final sample of likely GPs, explored in more detail in Section \ref{sec:res}	&	\S\ \ref{sec:final}   &	9628 (521 have non-SDSS redshifts, Table \ref{tab:nonSDSS GPs}, and 8313 are new, Table \ref{tab:GPcans}) \\
        \hline    
       7	&	Extended Peas	&	Subset of GP candidates (Sample 6) with half-light radii $>~1.1$"	&	Extended Peas are essentially absent among the currently known GPs, but may represent a merger (triggering) phase of GPs	&	\S\ \ref{sec:char}   &	$\sim$470\\
       \hline
    \end{tabular*}
    \caption{Description of various samples used in this work.}
    \label{tab:samples}
\end{table*}

\section{Principal Data Sources} \label{sec:Data Sources}

Our selection is based on the Sloan Digital Sky Survey (SDSS) Data Release 18 (DR18) photometric and spectroscopic data \citep{Almeida23}. We also use [OIII]5007 EWs from the MPA-JHU catalog \citep{Kauffmann03, Brinchmann04, Tremonti04}, as provided in the DR18 database (table \texttt{galSpecLine}). All EWs in this work refer to the rest-frame. As is the usual practice regarding EWs, no correction for internal dust is applied to it.

We combine the SDSS photometry with the photometry from the Wide-Field Infrared Survey Explorer (WISE; \citealt{Wright10}). Specifically, we utilize the unWISE catalog \citep{Lang16}, available from the SDSS database (table \texttt{wiseForcedTarget}), which provides photometry for 400 million sources where SDSS positions and light profiles served as modeling priors. 

We obtain SDSS and WISE data from the CasJobs SQL platform provided by SDSS via SciServer.   

\section{Green Pea Candidates Selection Methodology} \label{sec:ssmethods}

Our goal is to arrive at a relatively clean and complete sample of GP candidates using non-spectroscopic methods alone. To achieve this, we start with C09 color selection, but go beyond it in order to minimize substantial contamination from stars, quasars and artifacts. This contamination exceeds the number of genuine GPs by $\sim$20 times. We develop our method in three steps:

\begin{enumerate}
  \setlength\itemsep{0em}
    \item Introduce a cut to remove non-GP galaxies based on objects with known [OIII]5007 EW.
    \item Introduce three more cuts to remove non-GP contaminants based on objects with known identity (star, quasar, non-GP galaxy, GP).
    \item Remove most of the remaining non-GP contaminants and identify the final GP candidates using an SED matching technique.
\end{enumerate}

In developing the methodology we introduce different samples and sub-samples. Their description and purpose is summarized in Table \ref{tab:samples}. 

Before proceeding, it would be useful to define a GP galaxy for the purposes of this study. By GP we will consider a galaxy with strong [OIII]5007 emission (typically rest-frame EW $>100$ \AA), compact or unresolved appearance in SDSS images ($R_{50}<1.35$ arcsec) and lying in a redshift window that places the [OIII]5007 line (also calling it just [OIII]) in the $r$ band pass. GP-like objects are known in redshift ranges below and above this window \citep{Izotov11}, and have been occasionally referred to as `blueberries' and `purple grapes' \citep{Yang17,Brunker20}. Because the objects require different selection techniques, we do not consider them in this study.

\subsection{Removing Non-GP Galaxies from the Color Selected Sample} \label{sec: initial sample}

The first sample of GPs was established by C09. Their selection was constrained by a redshift range of $0.112 < z <0.36$, a requirement that the spectrum is classified as a galaxy, and color selection using four SDSS colors and five cuts: 

\begin{equation}
\begin{gathered}
u - r \leq 2.5\\
r - i \leq -0.2\\
r - z \leq 0.5\\
g - r \geq r - i + 0.5\\
u - r \geq 2.5(r - z)
\label{eq:c09}
\end{gathered}
\end{equation}

The redshift range used by C09 corresponds to the window wherein the [OIII]5007 line will be shifted into the $r$ band and result in broadband colors distinct from those of the majority of (but not all) stars, regular galaxies and quasars. This distinctness in color, especially in $r-i$ and $r-z$, is formalized in the above color cuts. The requirement that the spectrum is classified as a galaxy helps remove a small number of QSOs at these redshifts that may stray from the QSOs locus and make it through the color selection. Application of these criteria by C09, who used the SDSS Data Release 7 (DR7) spectroscopic sample, resulted in 251 objects.

We apply these C09 selection criteria onto the spectroscopic objects available from SDSS Data Release 18 (DR18), the most current data release as of the writing. The goal in this step is to find additional cut(s) to remove galaxies with lower [OIII] EW values, i.e., objects that are not GPs, and thus obtain a cleaner sample, as well as to confirm/refine the redshift window that results from C09 color selection. We perform our selection on the SDSS table \texttt{SpecPhoto}, requiring  galaxy spectra \texttt{class = `GALAXY'}. We apply the color cuts on valid, de-reddened model magnitudes (\texttt{modelMag}) with magnitude errors $<0.3$ in all bands (if not stated otherwise, the magnitudes we use throughout are model magnitudes de-reddened for the Milky Way dust extinction (but not for internal dust)). Note that the magnitude error cut amounts to a $z$-band selection, since the errors in other bands are almost always smaller than 0.3 mag. The average magnitude errors in ($ugriz$) are (0.12, 0.04, 0.03, 0.07, 0.17) mag. To be certain there are no GPs being missed, we initially allow objects in a slightly wider redshift range ($0.1<z<0.4$) than the one used by C09. 

The above criteria result in a sample of 1022 spectroscopic galaxies. Of those, 253 have valid [OIII]5007 EWs from the MPA/JHU catalog (See section \ref{sec:Data Sources}). The reason for such large incompleteness lies in the fact that the MPA/JHU catalog included in SDSS database (table \texttt{galSpecLine}) has not been updated based on new spectra obtained since SDSS DR8. Thus, the 253 objects with valid [OIII] EWs is practically the C09 sample (251 objects).

\begin{figure}[htbp]
    \centering
    \includegraphics[width=\columnwidth,clip=true,trim={8 0 0 0}]{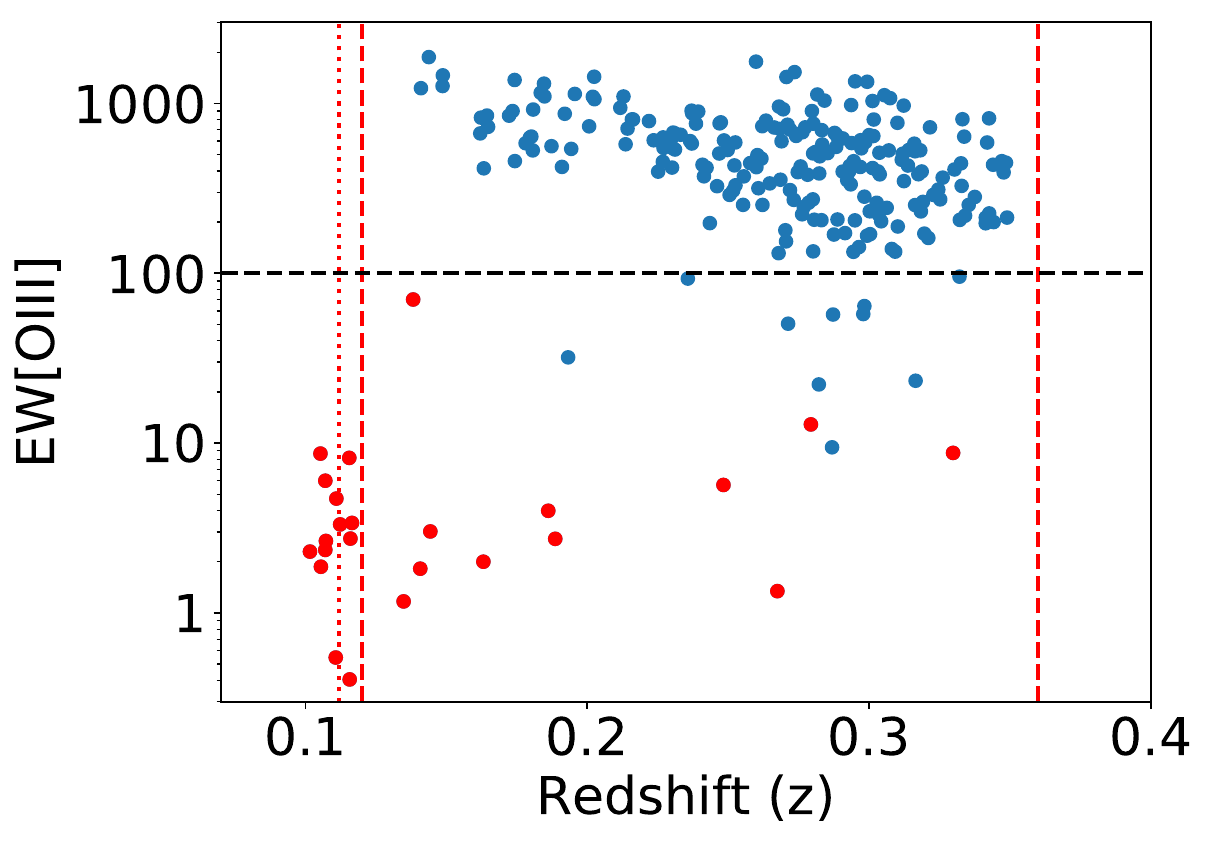}
    \includegraphics[width=\columnwidth,clip=true,trim={8 0 0 0}]{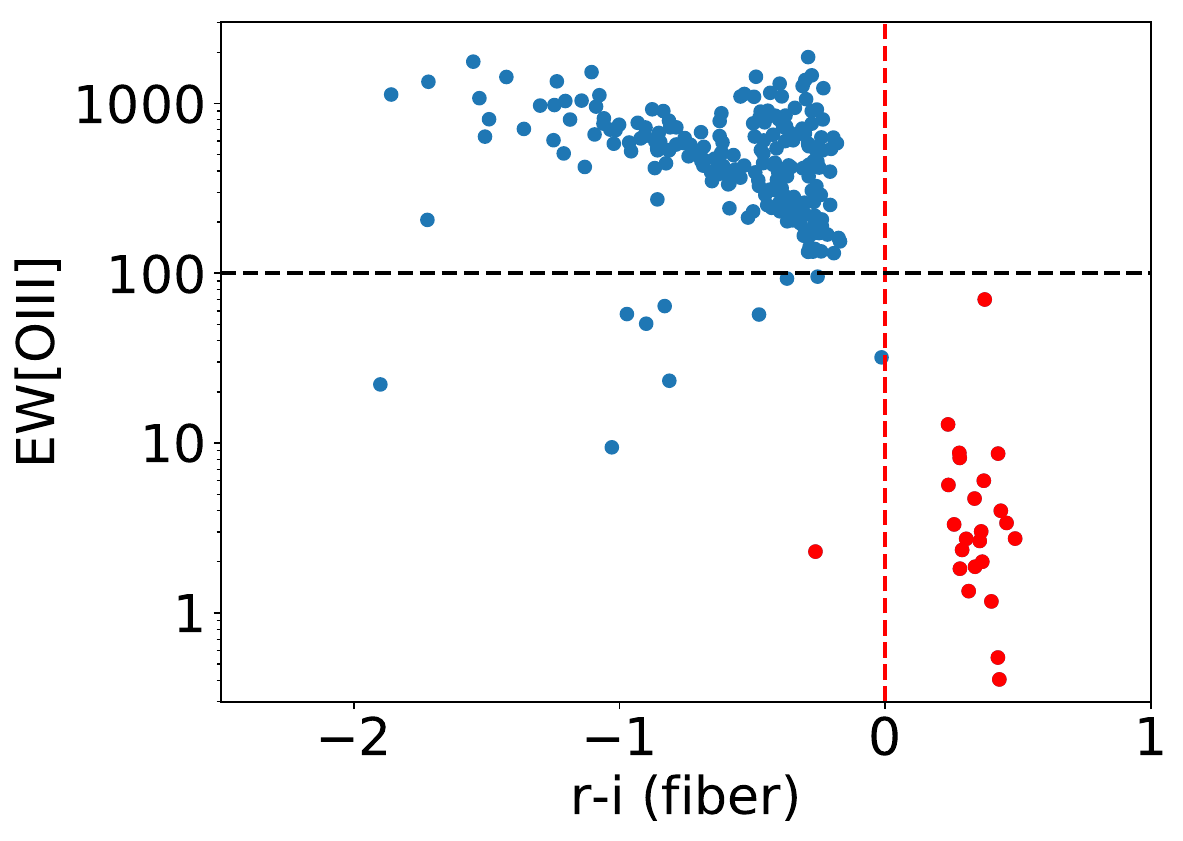}
    \caption{Removal of non-Green Pea interlopers from a \citet{Cardamone09}-like sample using [OIII]5007 equivalent width. Upper panel: EW([OIII]) vs.\ redshift plot allows us to refine the redshift window that yields GPs ($0.12<z<0.36$, red dashed lines). Lower Panel: EW([OIII]) vs.\ $(r-i)_{\mathrm{fiber}}$ color plot allows us to remove some non-GP interlopers (EW([OIII]) $<100$ \AA, below black dashed line) based on red color within the 3 arcsec (fiber) aperture ($(r-i)_{\mathrm{fiber}}>0$, red dashed line). Red dots denote the objects removed by either the fiber color cut or lying outside of the $0.12<z<0.36$ redshift window. All EWs in this work are in rest frame.}
    \label{fig: EWOIII vs z/rifiber}
\end{figure}

We first aim to refine the redshift range that selects GPs. The upper panel of \autoref{fig: EWOIII vs z/rifiber} shows [OIII]5007 EW against the redshift of the 253 galaxies. Of them, 39 have EW $<100$ Å, and therefore should not be considered GPs. The vertical dotted line in this plot denotes the C09 redshift cut at $z=0.112$. We see, however, that galaxies slightly above this cut,  up to $z=0.12$, all have low EWs. We have no low EW galaxies near $z=0.36$, so there is no need to modify C09's upper redshift limit. From here on, we will take $0.12< z <0.36$, represented by two vertical dashed lines, as the redshift window for which the color selection is effective.

The revised redshift limit does not remove all low-EW objects (see the upper panel of \autoref{fig: EWOIII vs z/rifiber}). Furthermore, the redshift cuts will obviously be of no use for selecting GPs when spectra are not available. We therefore look for potential ways to minimize non-GP contamination photometrically. In the lower panel of \autoref{fig: EWOIII vs z/rifiber}, we investigate the relationship between the [OIII]5007 EW and the $(r-i)_{\mathrm{fiber}}$, the de-reddened color (i.e., the color corrected for MW dust extinction) measured in the central 3” aperture corresponding to the spectroscopic fiber. Since GPs are compact ($R_{50}<1.35$ arcsec), there should not be much difference between the total $r-i$ colors used for the selection ($r - i \leq -0.2$), and the fiber $r-i$ colors. However, we see a distinct group of objects with red $(r-i)_{\mathrm{fiber}}$ colors in the lower right corner of the plot that do not have high enough EW to qualify as GPs under our definition (EW([OIII]) $>100$ Å). The discrepancy between their total colors and fiber colors suggests a more extended galaxy exhibiting a color gradient. Thus we introduce our first new cut, which requires $(r-i)_{\mathrm{fiber}}<0$. This cut removes 25 of 39 objects with EW([OIII]) $<100$ Å. All but one of these are classified as extended (\texttt{type = 3} in SDSS), and visually appear more like regular galaxies than GPs, thus justifying the exclusion. 

Our new cut based on fiber color is the first of several cuts that will help ensure compactness, which in turn reduces non-GP contamination.

\subsection{Removing Additional Contaminants Using Photometric Information} \label{sec: Prelim Sample}

We need to approach the situation of selecting GP candidates which will not have spectra/redshift information. Without being able to select objects in the redshift window, the color cuts alone will let in a large number of stars, quasars and non-GP galaxies. We are thus looking for photometric features that can reduce unwanted contaminants coming from redshift regimes outside of the GP redshift window ($0.12<z<0.36$). To identify these additional cuts we will utilize the fact that SDSS has obtained spectra for some of these contaminants. 

\begin{figure}[htbp]
    \centering
    \includegraphics[width=0.93\columnwidth,clip=true,trim={0 0 0 0}]{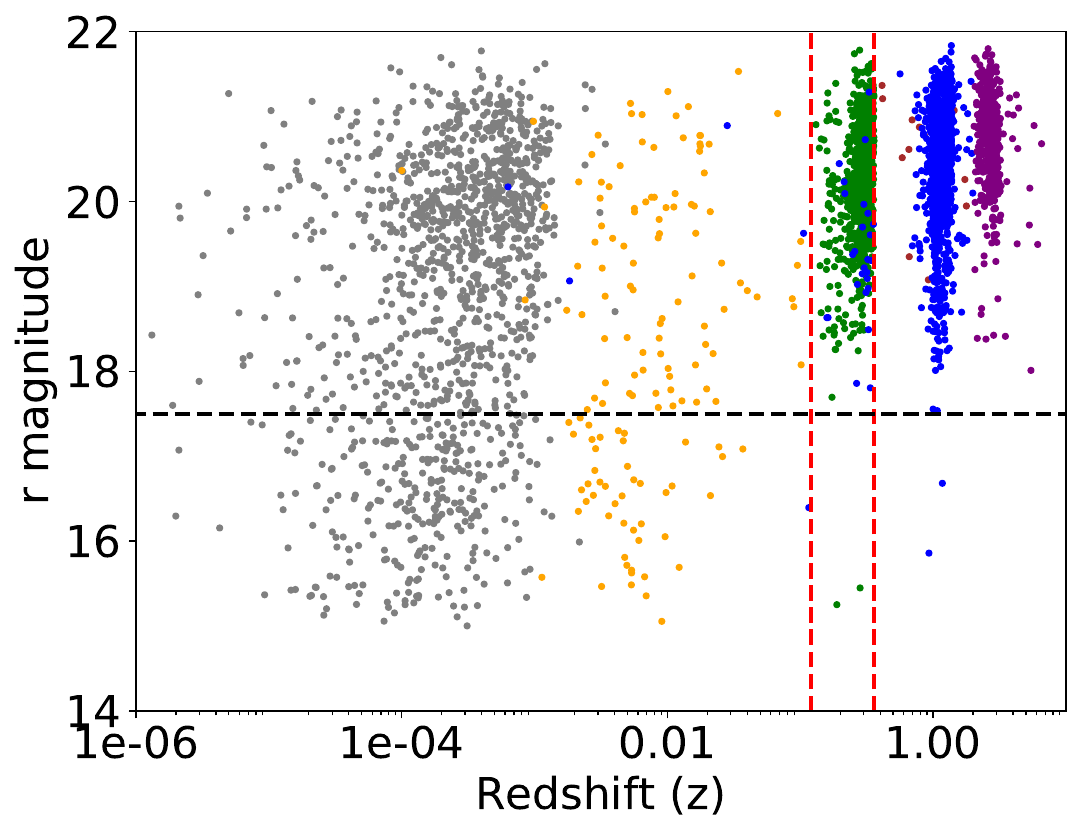}
    \includegraphics[width=0.93\columnwidth,clip=true,trim={0 0 0 0}]{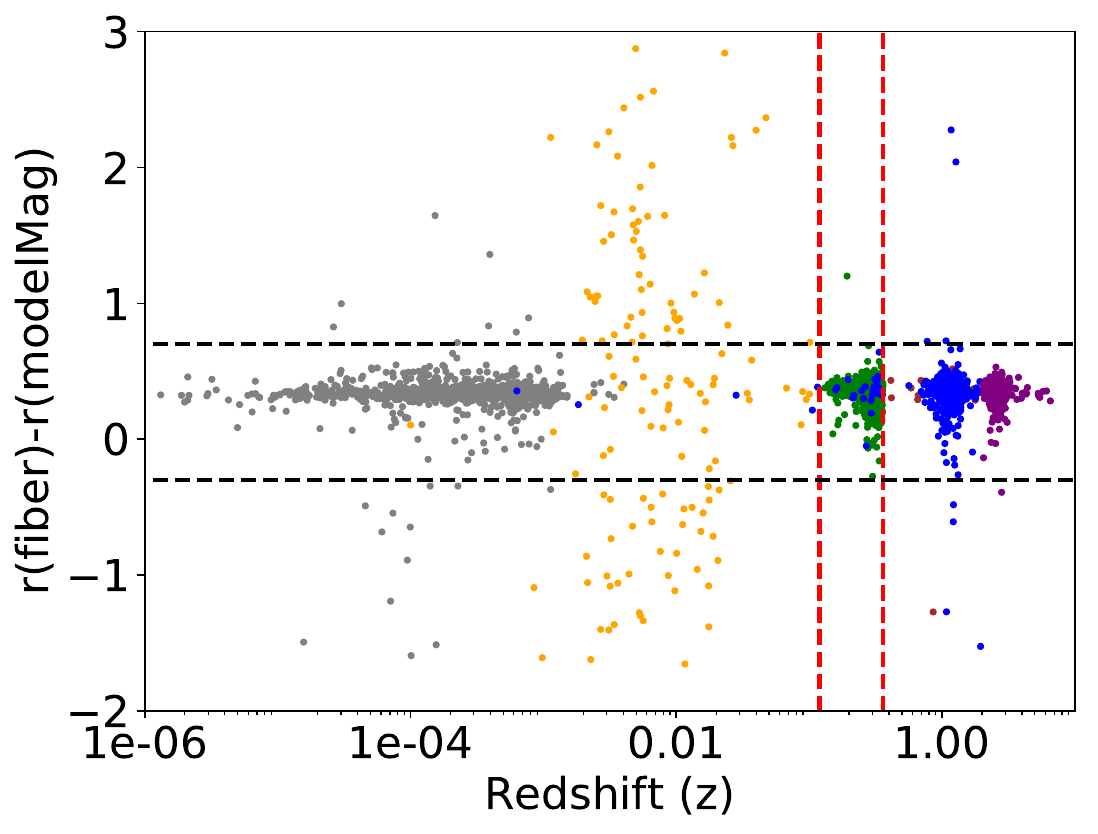}
    \includegraphics[width=0.93\columnwidth,clip=true,trim={0 0 0 0}]{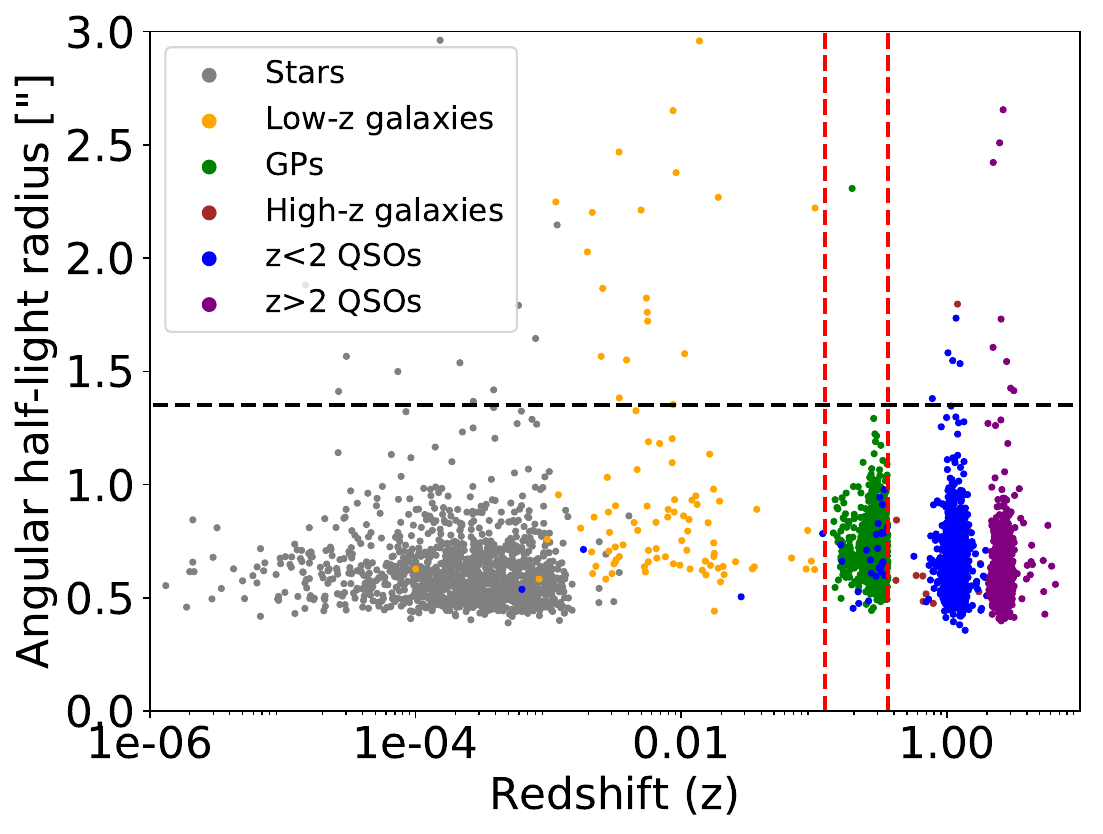}
    \caption{Removal of non-Green Pea contaminants using photometric information. Top: $r$ magnitude vs.\ redshift. Middle: fiber aperture (3") magnitude minus total magnitude ($r_{\mathrm{fiber}}-r_{\mathrm{model}}$) vs.\ redshift. Bottom: angular half-light radius in $r$ band ($R_{50}$) vs.\ redshift. Cuts in these three quantities (black dashed lines) allow us to remove additional non-GP contamination---mostly stars and low-$z$ galaxies or galaxy regions. The purpose of this cuts is to select a cleaner candidate pool for GPs when spectra are not available. Redshift is shown as the absolute value of redshift in order include the negative redshifts of many stars and nearby galaxies, which do not exceed $|z| = 0.004$. The redshift window ($0.12<z<0.36$) that yields GPs using the C09 color selection is shown with red dashed lines.}
    \label{fig: z vs cuts}
\end{figure}

We query the DR18 \texttt{SpecPhoto} table using the C09 color criteria, plus our new cut of $(r-i)_{\mathrm{fiber}} <0$. This query returns 4157 objects of any spectroscopic class and redshift. Using redshifts and SDSS spectroscopic classifications, we can separate objects into different categories: 
\begin{itemize}
  \setlength\itemsep{0em}
    \item Stars have \texttt{class =} \texttt{`STAR'}
    \item Low-redshift galaxies have \texttt{class =} \texttt{`GALAXY'} and $z<0.12$
    \item Green Peas (GPs) have \texttt{class =} \texttt{`GALAXY'} and $0.12 < z <0.36$
    \item High-redshift galaxies have \texttt{class =} \texttt{`GALAXY'} and $z>0.4$
    \item $z<2$ quasars have \texttt{class =} \texttt{`QSO'} and $z<2$
    \item $z>2$ quasars have \texttt{class =} \texttt{`QSO'} and $z>2$. 
\end{itemize}

\begin{figure*}[htbp]
    \centering
    \includegraphics[width=\textwidth]{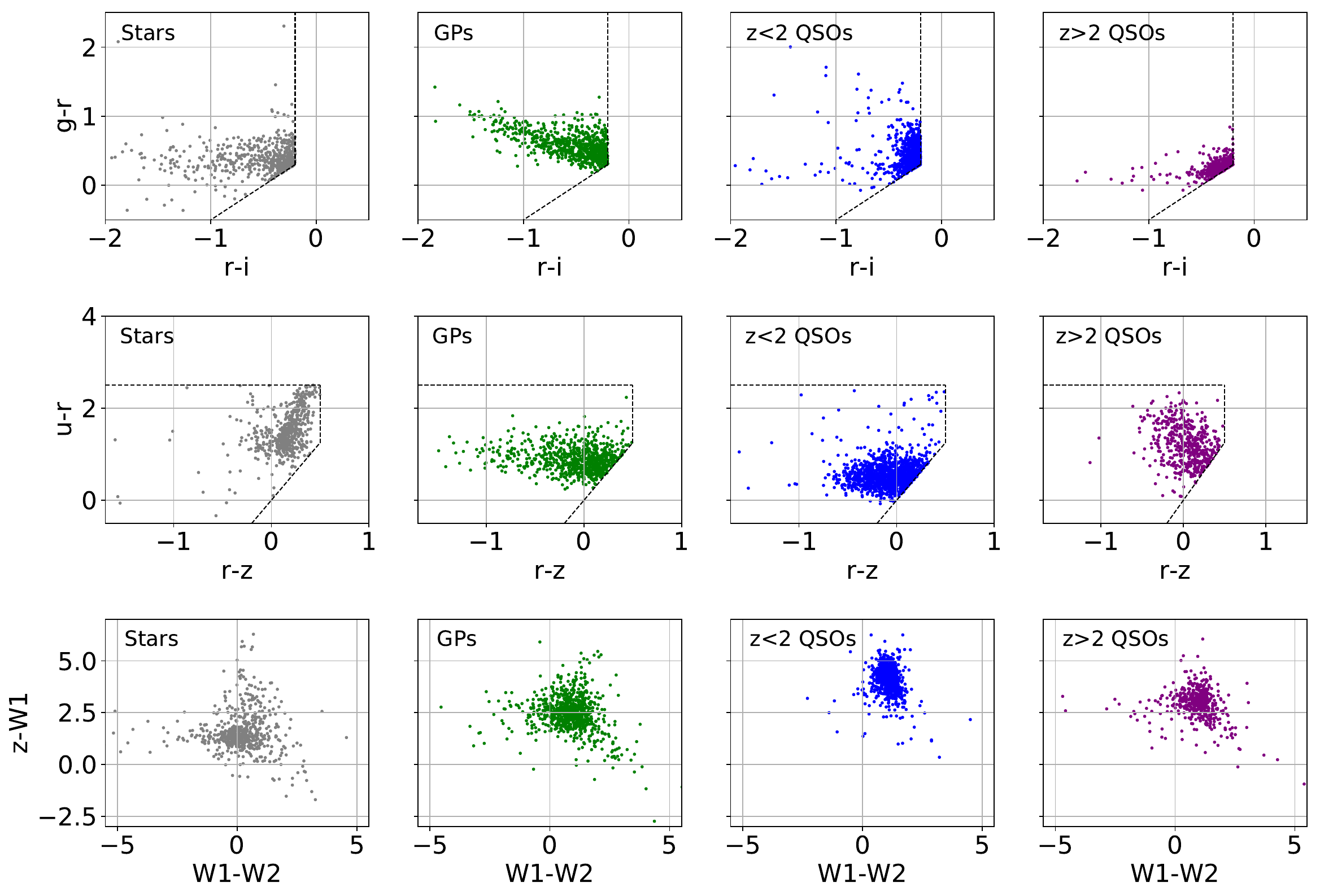}
    \caption{Three types of color-color diagrams showing Green Peas and the two main classes of contaminants (stars and quasars). Different classes of objects have somewhat distinct loci, but also overlap enough that no color selection can cleanly select just the GPs. \citet{Cardamone09} color cuts are shown as dashed lines.}
    \label{fig: known ccplts}
\end{figure*}

Contaminants will come from below the GP redshift limit ($z<0.12$) as stars, low-redshift galaxies and HII regions in nearby galaxies, while contaminants from above the limit ($z>0.36$) will mostly be from quasars. Thus, by investigating various photometric features against redshift, we hope to find criteria to help remove some of the contaminants, while preserving the GPs. 

The top panel in \autoref{fig: z vs cuts} shows $r$ magnitude versus the redshift for our 4157 objects. The red dashed lines represent our redshift range for GPs, $0.12< z <0.36$. We see that there are essentially no GPs brighter than $r=17.5$. We thus introduce a cut at $r>17.5$ to exclude bright stars and low-redshift galaxies or HII regions. This cut removes 350 objects.

We next find that we can remove extended objects in two ways: require consistent total and aperture (fiber) magnitudes (middle panel in \autoref{fig: z vs cuts}) as well as require small sizes (bottom panel in \autoref{fig: z vs cuts}). By requiring $-0.3< r_{\mathrm{fiber}}-r_{\mathrm{model}}<0.7$, which leaves GPs intact (black dashed lines in the middle panel of \autoref{fig: z vs cuts}), we are removing objects typically due to bursty regions in large $z\lesssim0.03$ galaxies, detected in SDSS as individual sources. This cut removes 140 objects. To account for GPs' compact size, we implement another cut by requiring $R_{50}<1.35$ arcsec (dashed line in the bottom panel of \autoref{fig: z vs cuts}, where $R_{50}$ is the radius that contains 50\% of the Petrosian flux in the $r$ band (\texttt{petroR50\_r}). This cut removes 99 objects.

To summarize, based on the analysis in Section \ref{sec: initial sample} and this section, we have identified a total of four selection cuts that help remove some of the non-GP contaminants (mostly bright stars and extended objects):

\begin{equation}
\begin{gathered}
(r-i)_{\mathrm{fiber}} < 0 \\
r\, (\equiv r_{\mathrm{model}}) > 17.5 \\
-0.3 < r_{\mathrm{fiber}} - r_{\mathrm{model}} < 0.7 \\
R_{50} < 1\farcs 35
\label{eq:crit}
\end{gathered}
\end{equation}

\noindent We experimented basing the above cuts on various other colors and sizes, but found these to be most effective. Overall, these new cuts are a useful expansion upon the original C09 color cuts (Eq.\ \ref{eq:c09}) , whether the spectra are available or not, and especially when they are not, as is the goal in this work. To illustrate this, we note that the number of \textit{photometric} objects (among which we hope to identify new GPs) selected using only the original color-selection criteria (Eq.\ \ref{eq:c09}) is 184,002 (i.e., roughly 200$\times$ more than the spectroscopically confirmed GPs). However, with the application of the new cuts (Eq.\ \ref{eq:crit}), this number is reduced by 68\%, leaving 58,761 objects, thus removing a large fraction of contaminants. 

\subsection{Contaminant Removal Using SED Matching Method} \label{sec: Refined sample}

Even with the refined photometric selection criteria for GPs, it is difficult to always distinguish them from the contaminants that are similarly compact and of similar brightness (see Figure \ref{fig: z vs cuts}). To establish if there is additional photometric information that can be exploited to separate GPs from stars and quasars, we explore various color-color diagrams. For this we use 3133 objects of any spectroscopic class and redshift (whose nature is thus known from their spectra), which pass the C09 color selection plus the four additional cuts introduced in Sections \ref{sec: initial sample} and \ref{sec: Prelim Sample} and are detected in WISE bands W1 and W2. The need for WISE photometry will be explained shortly. We call this the reference set. It includes genuine GPs as well as their contaminants (mostly stars and quasars).

The top row of \autoref{fig: known ccplts} displays the $g-r$ vs.\ $r-i$ color-color plots for GPs, and two main classes of contaminants: stars and quasars. The GPs have a noticeable upwards spur, which is also visible in the left panel of C09 Fig.\ 2. Stars, on the other hand, extend more horizontally (constant $g-r$), but the overlap is significant between these two categories. Quasars (both $z<2$ and $z>2$) also overlap with the red portion of GP distribution (in $r-i$).  The middle row of \autoref{fig: known ccplts} features these categories in $u-r$ vs.\ $r-z$ space, where we again see substantial overlaps, especially between GPs and $z<2$ quasars. The bottom row looks at $z-$W1 vs.\ W1$-$W2 space, where W1 and W2 are 3.4 and 4.6 $\mu$m bands from WISE.  We see that when WISE colors are included the GPs and $z<2$ quasars are more distinctly distributed. However, the overlaps are still significant. Apparently, we cannot use simple color-color cuts to cleanly separate GPs from contaminants.

Instead, our last step for the selection of GPs proceeds by using all of the optical and near-IR photometry simultaneously---what we refer to as an SED \textit{matching} method. This method will compare every object from the photometric candidate pool with each and every object of known classification (the reference set), in order to determine which classification (i.e., star, GP, QSO, etc.) is the best match. We do this by utilizing $r$ magnitude and $u-g$, $g-r$, $r-i$, $i-z$, $z-$W1, and W1$-$W2 colors to create a metric: 
\begin{align}
D^2 = {} & (\Delta r)^2 + [\Delta (u-g)]^2 + [\Delta (g-r)]^2 \notag \\
        & + [\Delta (r-i)]^2 + [\Delta (i-z)]^2 \notag \\
        & + [\Delta (z-\mathrm{W1})]^2 + [\Delta (\mathrm{W1}-\mathrm{W2})]^2
\label{eq:d2}
\end{align}

\noindent where $\Delta$ is the difference in a given quantity between the candidate and the reference set objects. For each object from the candidate pool, we will calculate $D^2$ relative to all 3133 reference objects. The closest match (i.e., the reference object that returns the smallest $D^2$) will determine the classification of the candidate object. This method is similar to the SED fitting method, except that the comparison is (a) against the actual, rather than the model, photometry, (b) includes both colors and a magnitude and (c) does not weight the terms by their error. We include the magnitude in the metric because the overall brightness provides additional constraining information different from the relative fluxes (colors). For example, a bright candidate might be more likely to be a star than a distant quasar. We do not weight the terms by their errors in order not to bias the metric towards the brighter reference objects.

\begin{deluxetable}{cccc}
\tablecaption{Composition of the reference set and the results of self-validation test.  
\label{tab:Ref classes}}        
\tablewidth{0pt}                                                  
\tablehead{
\colhead{Class} & 
\colhead{\rule{0pt}{20pt}\shortstack{Share}} & 
\colhead{\shortstack{Assigned class of \\ actual GPs}}
}
\startdata                                                            
Green Peas & 26\% & 84\%\\                           
$z<2$ QSOs & 39\% & 6\%\\                           
$z>2$ QSOs & 16\% & 6\%\\                           
Stars & 19\% & 3\%\\                           
Low-$z$ galaxies & 0.2\% & 0.1\%\\                           
High-$z$ galaxies & 0.7\% & 0.5\%\\                                     
\enddata
\tablecomments{The reference set consists of all objects with spectra that pass the color cuts (Table \ref{tab:samples}. Categories of different objects in the reference set are in Column 2. Category assigned by the SED matching classification of the actual 799 GPs in the reference set (self-validation text) are in Column 3.}
\end{deluxetable}

Before running the SED matching method on a candidate pool, we first perform self-validation, by comparing the reference set to itself, but, of course, not allowing a given object to match itself. This will produce a ``reclassification" of objects with known classifications, so we can evaluate the reliability of our method and its purity.

\begin{figure*}[htbp]
    \centering
     \includegraphics[width=1.05\columnwidth,clip=true,trim={18 10 110 30}]{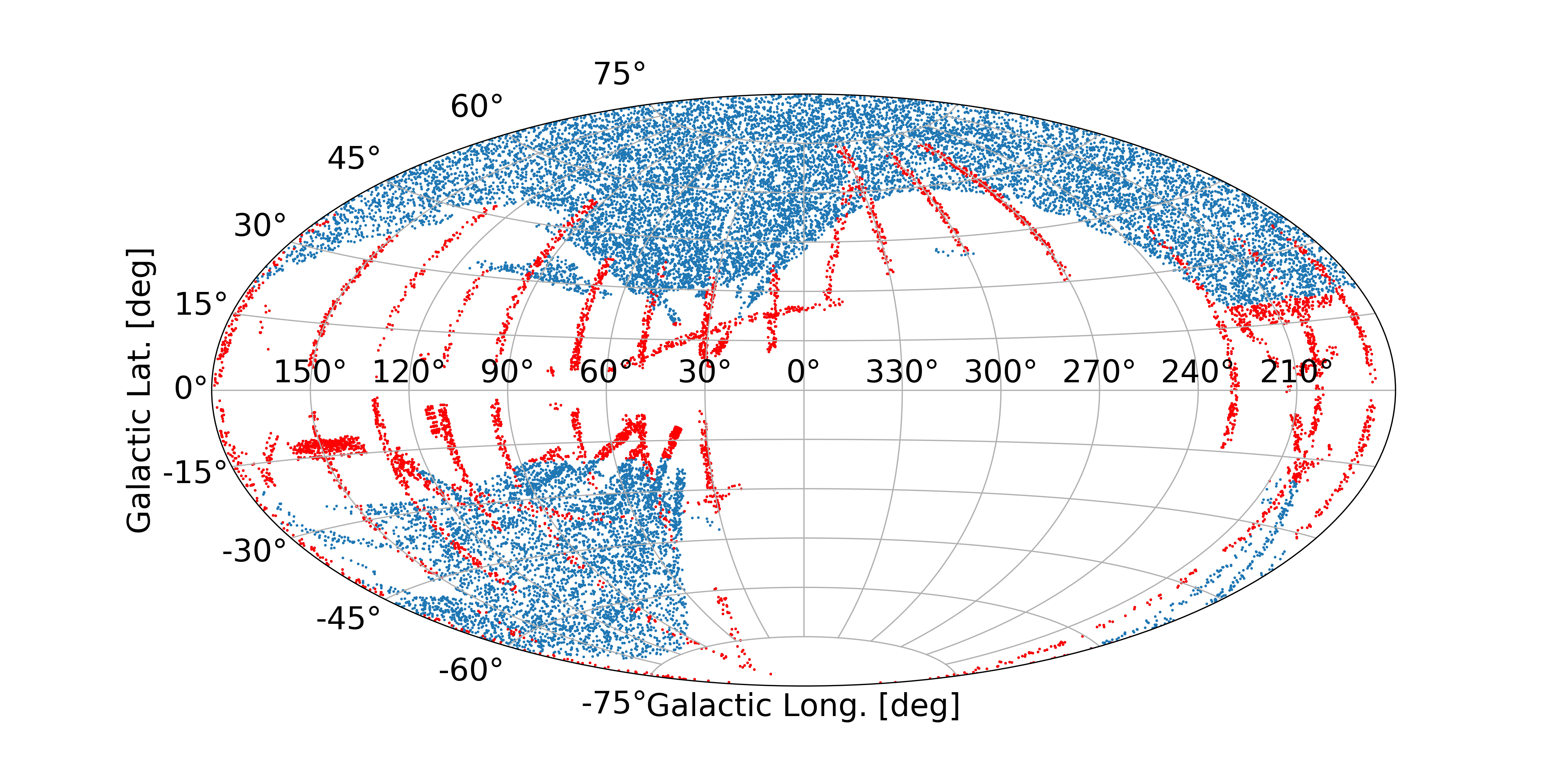}
    \includegraphics[width=1.05\columnwidth,clip=true,trim={18 10 110 30}]{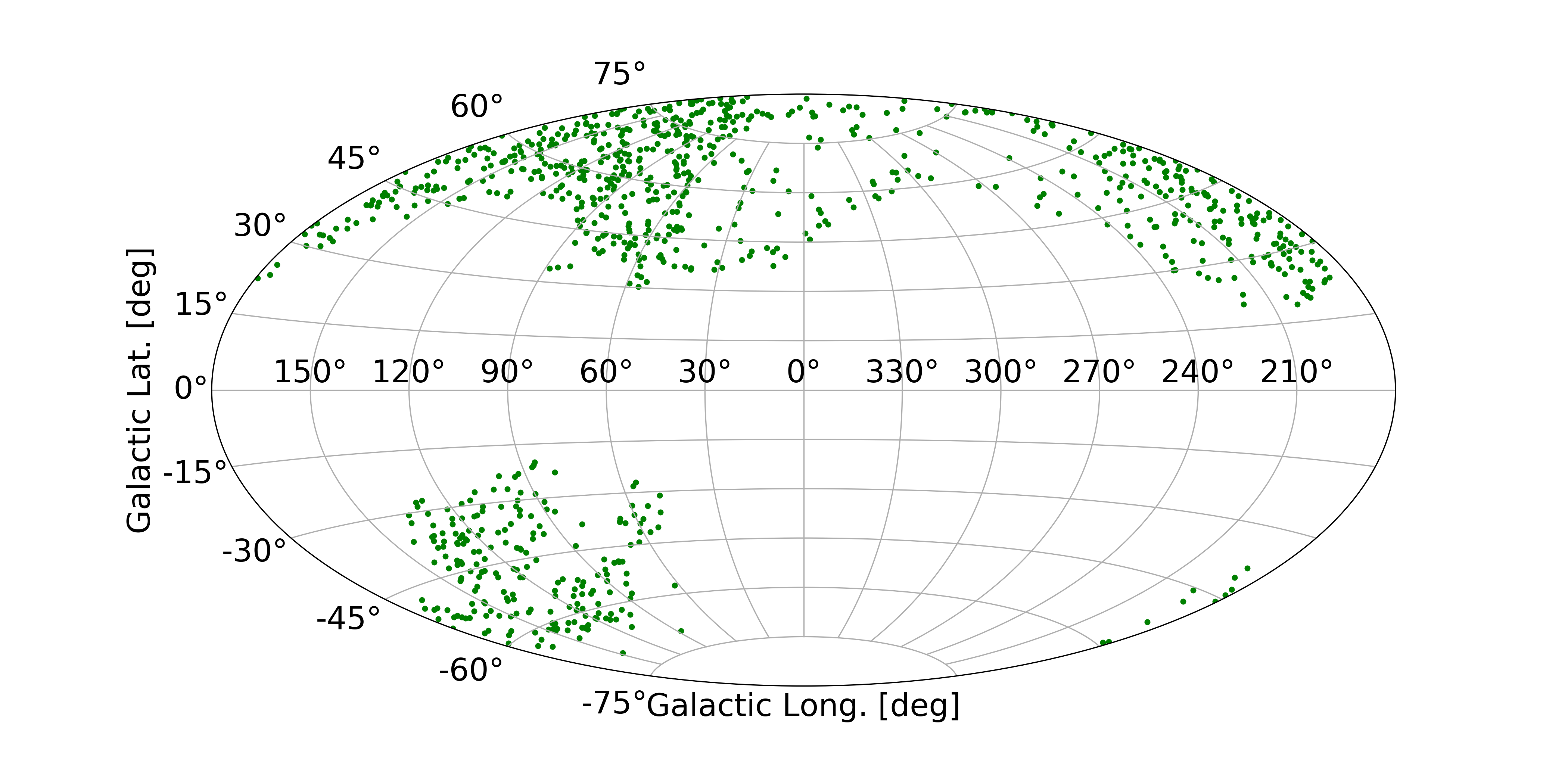}
    \caption{Galactic maps showing the Green Pea candidate pool (left) and spectroscopically confirmed Green Peas (right). Removing the objects plotted in red from the candidate pool (low latitude objects and objects in isolated SEGUE stripes) helps reduce the stellar contamination and provides a more contiguous area.}
    \label{fig: Positional cut galactic}
\end{figure*}

In the reference set, 799 objects are known to be GPs. SED matching classifies 670 of them as Green Peas. This means the method has 84\% efficiency (recovery rate) at identifying GPs. We see from Table \ref{tab:Ref classes} that a true GP is most often mistaken for a QSO. Using the SED matching method but without the WISE terms in Eq.\ \ref{eq:d2} returns an efficiency of recovery of 75\%, so the SED matching is not as effective when using only $ugriz$ photometry. The recovery rate is not expected to be uniform. We find it to be 77\% for the less extreme GPs (EW[OIII] $< 300$ \AA), rising to 94\% for the more extreme ones (EW[OIII] $> 700$ \AA). This is because the lower EW GPs tend to be in the regions of photometric space where contaminants are more common.

We tested changing the relative weight of the magnitude term with respect to the color terms, but that did not result in any improvements in the rate of recovery. Note that the efficiency established here addresses how well we can recover the GPs that would be selected using our cuts and SDSS/WISE survey depths. Testing the overall completeness with respect to a full population of GPs is discussed in Section \ref{sec:disc}.

From the self-matching we can also get an approximate assessment of the contamination rate. Our classification places 786 objects into the GP category, of which 15\% in reality are not GPs. As we will see in Section \ref{sec:final}, the more accurate assessment of the contamination rate appears to be lower. 

After investigating the effectiveness of the SED matching method and finding it to be relatively high, we now want to apply it to objects without spectra, in order to identify new GP candidates.

\subsection{Candidate Pool and the Selection of Green Pea Candidates} \label{sec:final}

With the methodology for identifying likely GPs fully developed, we now select the candidate pool---photometric objects that mostly do not have spectra---and perform the SED matching against the objects with spectra, and therefore, of known classification (the reference objects). 

As stated previously, there are 184,002 DR18 photometric objects that pass C09 color cuts, of which 58,761 further pass our four new cuts (fiber color cut, magnitude cut, size cut and fiber vs.\ total magnitude consistency cut). To apply the SED matching method, we also require valid W1 and W2 magnitudes, which brings us to 40,187 sources. Requiring WISE detections automatically removes any SDSS artifact sources (e.g., cosmic rays), since they will not appear in WISE. Looking at the sky distribution of these 40,187 sources (\autoref{fig: Positional cut galactic}, left panel) we see that unlike the spectroscopically confirmed GPs that come from contiguous, high-latitude regions (\autoref{fig: Positional cut galactic}, right panel), many photometric sources are located in stripes.  The stripes come from the SEGUE survey that covers areas close to the Galactic plane and is thus dominated by stars, thus potentially exacerbating contamination. To remove objects from SEGUE stripes we require \texttt{stripe} $\leq 86$. We also remove any remaining objects within 20 degrees of the plane, which will also be star dominated. This brings us to the final candidate pool of 18,880 objects. The reduction in the candidate pool (53\%) is disproportionate to the area that was removed (red dots in \autoref{fig: Positional cut galactic}, left panel) most likely because of the high concentration of stars in these areas compared to the rest of the area. The SQL query that includes all the criteria for the final candidate pool is given in Appendix \ref{app}.

Note that we do not remove SEGUE/low-latitude objects from the spectroscopic reference set. There are only 6\% of the reference sample in these areas. Having a few more reference objects (even if they are predominantly stars) does not affect our ability to classify the candidate pool (for which we remove these star-dominated regions) and does not affect our final contamination rate estimate, since it will take the difference in the composition of the candidate pool and the reference set into account.

\begin{figure*}[htbp]
    \centering
    \includegraphics[width=0.3\textwidth]{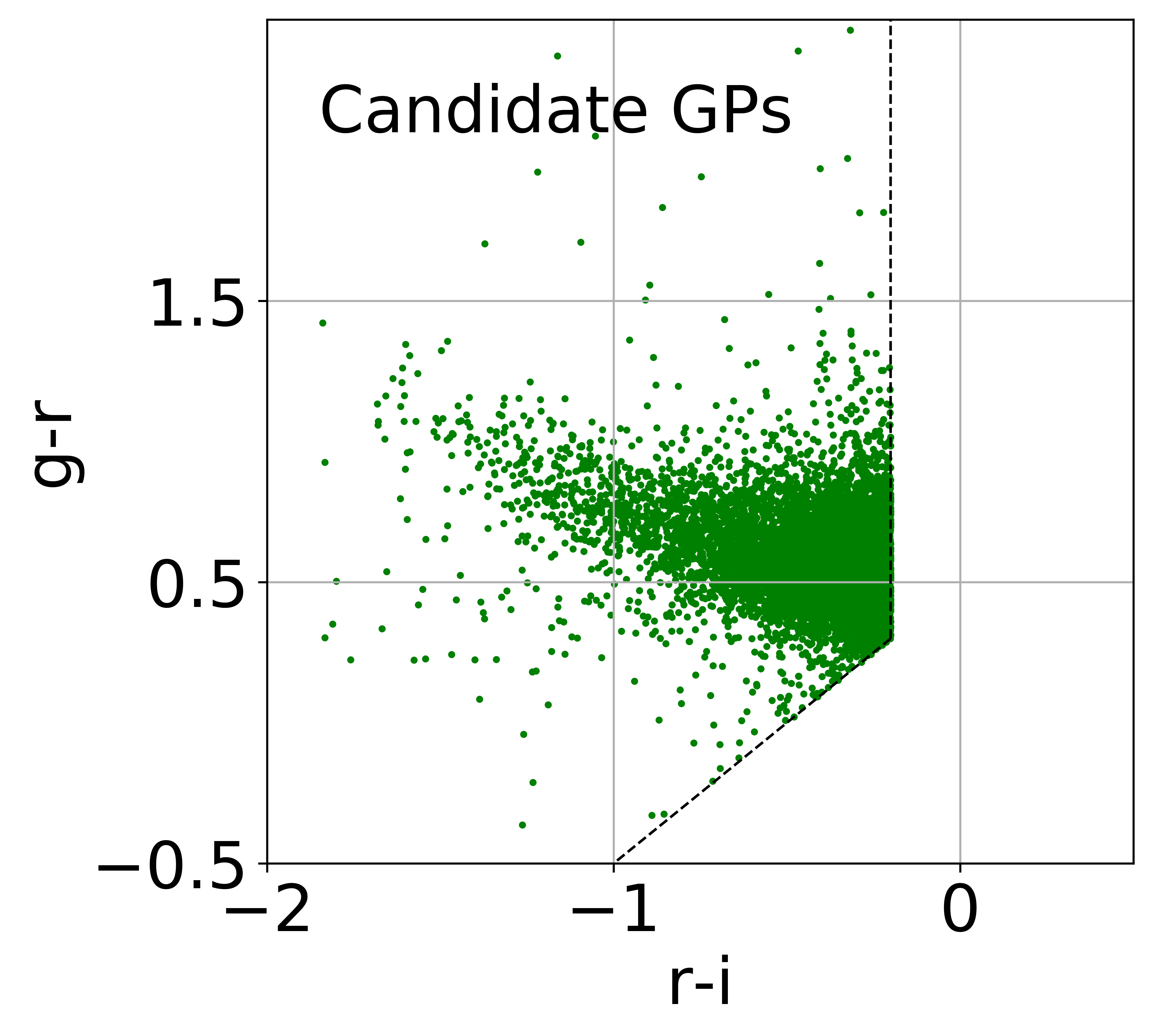}
    \includegraphics[width=0.3\textwidth]{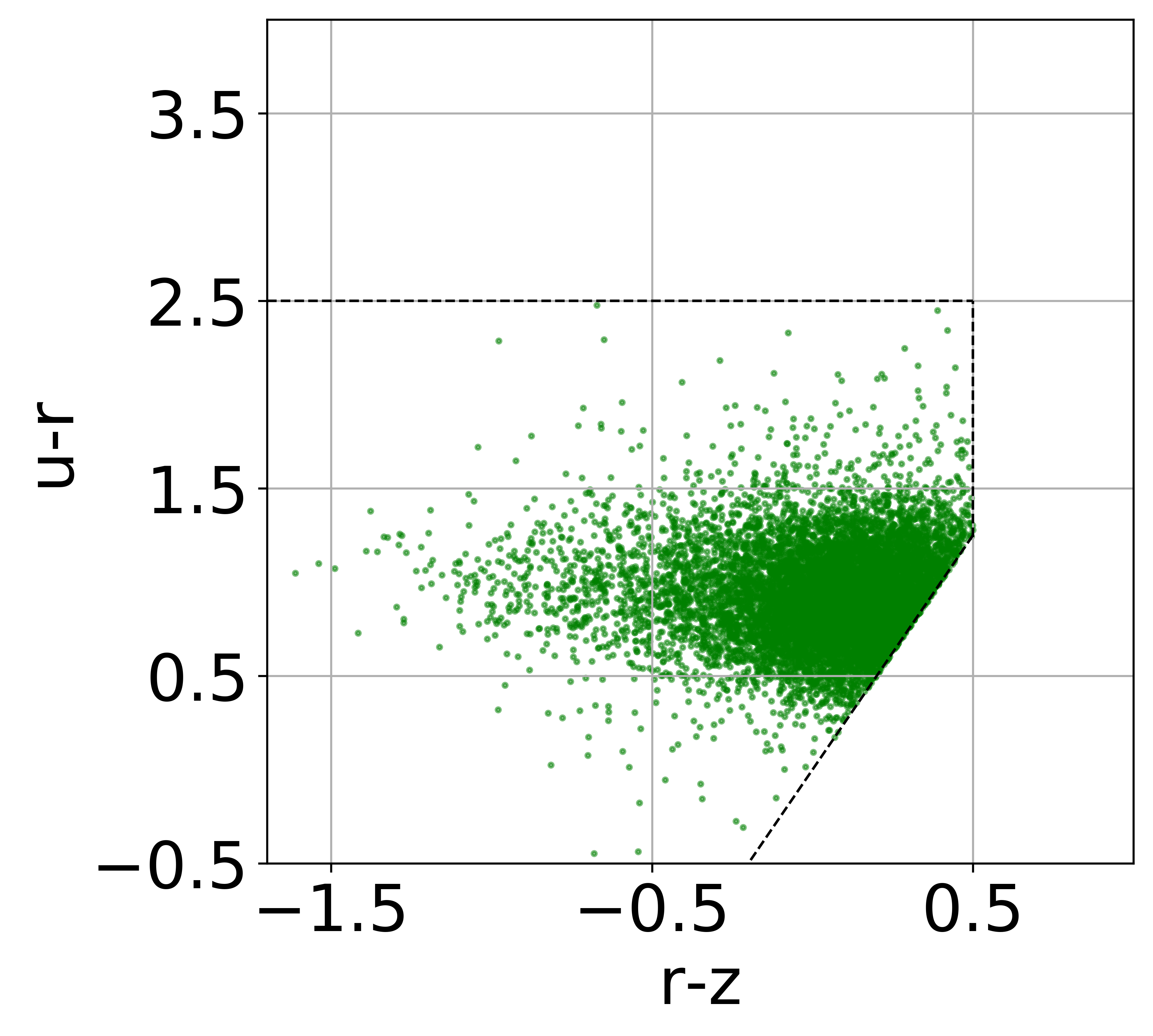}
    \includegraphics[width=0.3\textwidth]{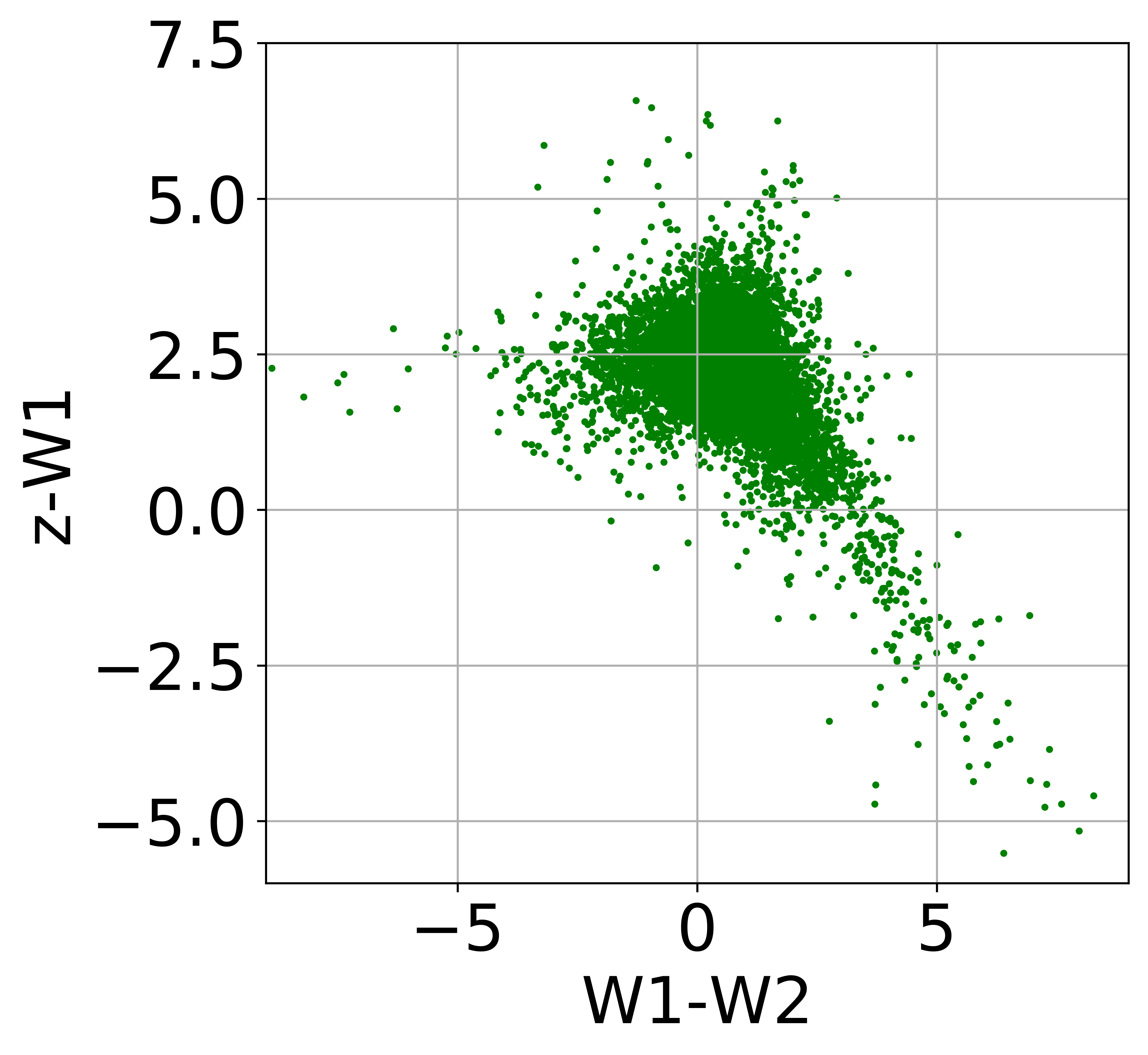}
    \caption{Color-color plots of Green Pea candidates identified using the SED matching method. Left: $g-r$ vs.\ r-i color-color plot. Middle: u-r vs.\ r-z color-color plot. Right: z-W1 vs.\ W1-W2 color-color plot. Distribution of GP candidates matches that of the spectroscopically confirmed GPs (second column of Fig. \ref{fig: known ccplts}). Dashed lines indicate \citet{Cardamone09} selection cuts.}
    \label{fig: GP cans ccplts}
\end{figure*}

As a final step, we subject the candidate pool to the SED matching with respect to the reference objects. We get 9628 objects (51\% of the pool) classified as GPs. This is some 10 times more than the number of GPs that can be spectroscopically selected in DR18 (Section \ref{sec: initial sample}) and 44 times more than the number of spectroscopically selected GPs in DR7.  Numbers pertaining to other categories that the candidate pool objects get classified into are listed in Table \ref{tab:GPcans class}.

\begin{deluxetable}{cccc}
\tablecaption{Classification of the Candidate Pool Objects
\label{tab:GPcans class}}        
\tablewidth{0pt}                                                  
\tablehead{
\colhead{\rule{0pt}{10pt} Classified as} & 
\colhead{Count} &
\colhead{Share}
}
\startdata                                                            
Green Peas & 9628 & 51\% &  \\                           
$z<2$ QSOs & 3403 & 18\% &  \\                           
$z>2$ QSOs & 1742 & 9\% &  \\                           
Stars & 4004 & 21\% &  \\                           
Low-$z$ galaxies & 29 & 0.2\% &  \\                          
High-$z$ galaxies & 74 & 0.4\% &  \\                                     
\enddata
\end{deluxetable}

Figure \ref{fig: GP cans ccplts} features our GP candidates in three different color-color spaces. We notice similar features to the spectroscopically confirmed GPs shown in Figure \ref{fig: known ccplts}, such as the  upward spur in $g-r$ vs.\ $r-i$. The comparison gives us confidence that the method worked as intended.

In Section \ref{sec: Refined sample}, we estimated, from self-validation on the reference set, that the contamination rate of the candidate GPs is 15\%. However, the actual contamination rate will depend on how common the contaminant categories are among the candidates, which need not be the same as in the reference set. Namely, the spectroscopic selection in SDSS may have favored one type of objects over another. Indeed, quasars, which accounted for the bulk of the contamination in the reference set, are actually only half as common in the candidate pool (compare the shares in Tables \ref{tab:Ref classes} and \ref{tab:GPcans class}), so they will contribute half as much as contaminants. This is presumably because quasars were more intentionally targeted by SDSS in this color parameter space. The share of stars is similar, so their contribution to contamination will stay the same. These adjustments result in a 10\% contamination rate for our GP candidates, equally distributed among $z<2$ QSOs, $z>2$ QSOs, and stars. Note that the definitive determination of the purity (contamination rate), requires a spectroscopic follow up of the candidates.

\begin{figure}[htbp]
    \centering
    \includegraphics[width=0.8\columnwidth,clip=true,trim={8 0 10 8}]{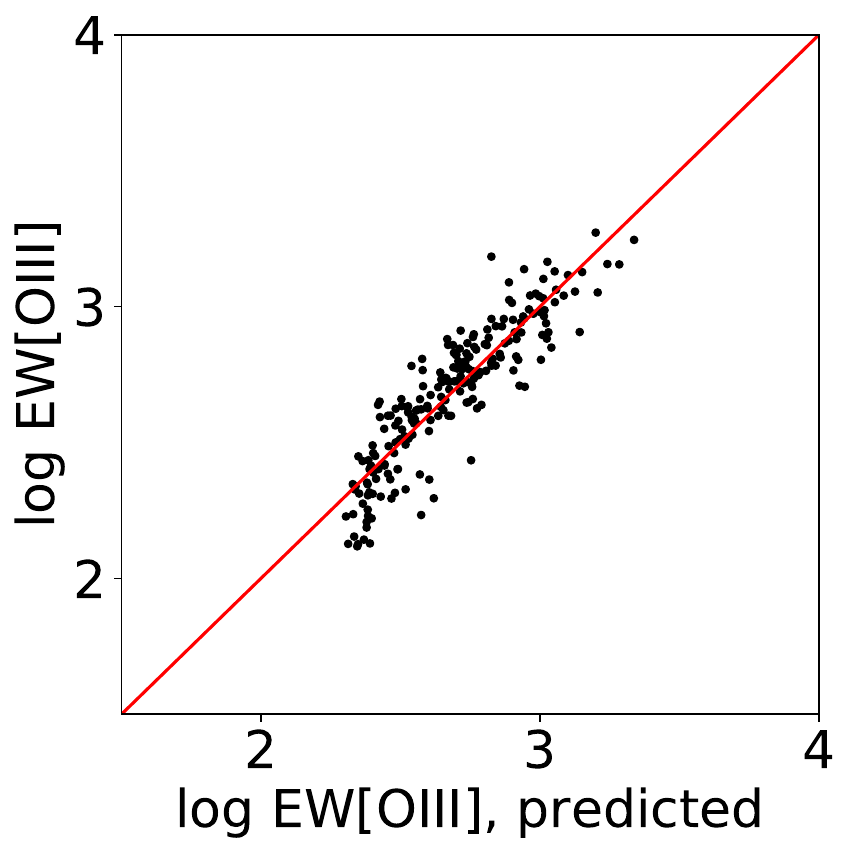}
    \caption{Comparison of the actual [OIII]5007 equivalent widths (EW) of Green Peas to the equivalent widths predicted from broad-band colors. Predicted EW is based on a combination of $g-r$, $r-i$ and $r-z$ colors (Eq. \ref{eq:ew}). The scatter is only 0.11 dex. Based on 214 GPs with [OIII] EWs from the MPA/JHU catalog.}
    \label{fig: ewoiii vs pred}
\end{figure}

\begin{figure*}[htbp]
    \centering
    \includegraphics[width=0.9\textwidth,clip=true,trim={0 0 0 0}]{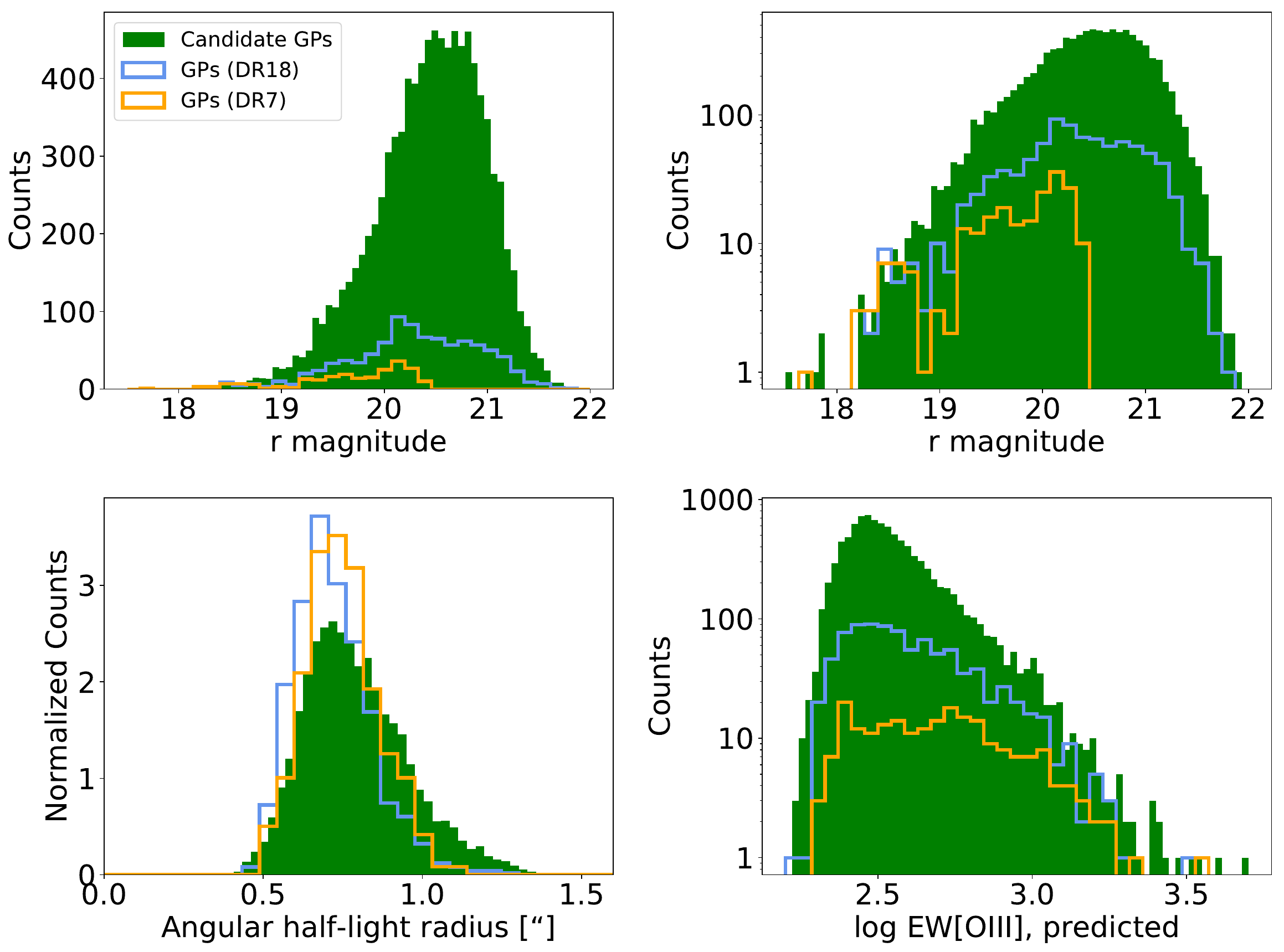}
    \caption{Distribution of $r$-band magnitudes (top left, linear scale; top right, log scale), half-light radii ($R_{50}$, lower left) and estimated [OIII]5007 equivalent widths (lower right) for Green Pea candidates (green solid histograms) and spectroscopically confirmed Green Peas (blue presents current (DR18) sample and orange are GPs that were available to C09 (DR7)). Distributions of half-light radii are normalized. New GP candidates are much more numerous than the spectroscopically confirmed ones and extend to somewhat larger sizes.}
    \label{fig: GPcans characteristics}
\end{figure*}

\section{Characterization of Green Pea Candidates} \label{sec:res}

\subsection{Estimation of [OIII]5007 \AA\ Equivalent Width from Broad-band Colors} \label{sec:ew}

In what follows, it would be useful to estimate how extreme the new GP candidates are in terms of the [OIII]5007 equivalent width, even though spectra are not available. Such information will also be important for selecting the most promising GP candidates for follow up.

\begin{figure*}[htbp]
    \centering
    \includegraphics[width=0.18\textwidth,height=0.18\textheight,keepaspectratio,clip=true]{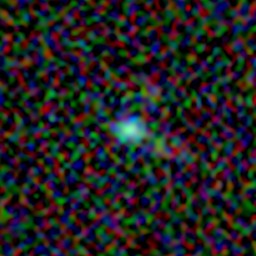}
    \includegraphics[width=0.18\textwidth,height=0.18\textheight,keepaspectratio,clip=true]{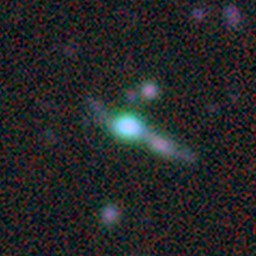}
    \hspace{0.05\textwidth}
    \includegraphics[width=0.18\textwidth,height=0.18\textheight,keepaspectratio,clip=true]{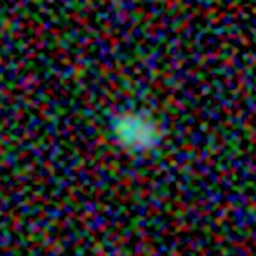}
    \includegraphics[width=0.18\textwidth,height=0.18\textheight,keepaspectratio,clip=true]{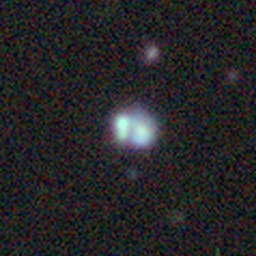} \\
    \vspace{0.01\textwidth}
    \includegraphics[width=0.18\textwidth,height=0.18\textheight,keepaspectratio,clip=true]{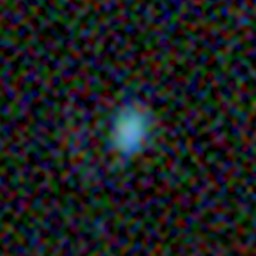}
    \includegraphics[width=0.18\textwidth,height=0.18\textheight,keepaspectratio,clip=true]{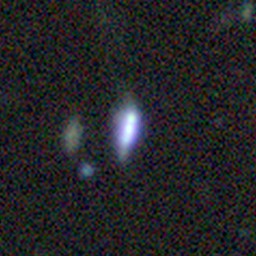}
    \hspace{0.05\textwidth}
    \includegraphics[width=0.18\textwidth,height=0.18\textheight,keepaspectratio,clip=true]{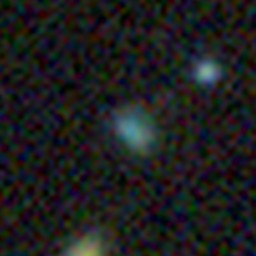}
    \includegraphics[width=0.18\textwidth,height=0.18\textheight,keepaspectratio,clip=true]{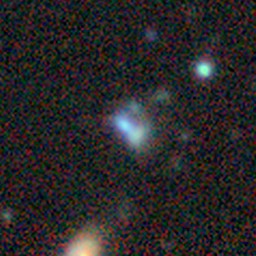} \\
    \vspace{0.01\textwidth}
    \includegraphics[width=0.18\textwidth,height=0.18\textheight,keepaspectratio,clip=true]{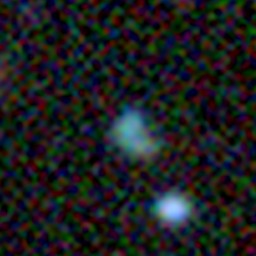}
    \includegraphics[width=0.18\textwidth,height=0.18\textheight,keepaspectratio,clip=true]{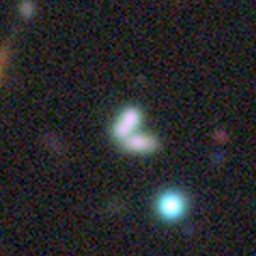}
    \hspace{0.05\textwidth}
    \includegraphics[width=0.18\textwidth,height=0.18\textheight,keepaspectratio,clip=true]{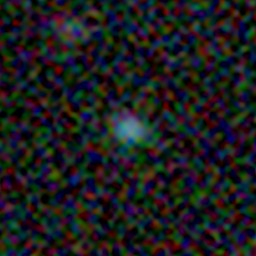}
    \includegraphics[width=0.18\textwidth,height=0.18\textheight,keepaspectratio,clip=true]{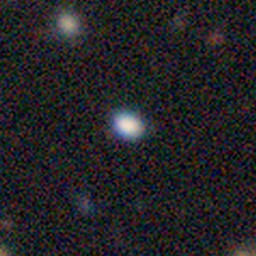}
    \caption{Examples of the ``Extended Peas". Shown are 6 out of 41 GP candidates with $R_{50}>1.1$ arcsec that have been imaged in Subaru HSC DR3. Left columns in each of the two stacks are SDSS images (with typical seeing of 1.3 arcsec), and right columns are Subaru images (typical seeing 0.7 arcsec). Extended Peas are rare among the known (spectroscopically confirmed) GPs. Many of the Extended Peas appear in Subaru images as mergers or as close pairs, which is less obvious in SDSS images. Each cutout is 20 arcsec on a side. Cutouts were obtained using the Legacy Survey Sky Viewer (\url{https://www.legacysurvey.org/viewer}).}
    \label{fig:Extended GPs}
\end{figure*}

Here, we investigate if the broadband colors can be used to estimate the EW of the [OIII]5007 line in GPs. We already expect that $r-i$ or $r-z$ individually should correlate with the [OIII] EW, but we hope that using more than one color would provide a tighter correlation, because, for example, it might mitigate against the fact that the colors are subject to redshifting. 

To do so, we use 214 spectroscopically confirmed GPs that have [OIII]5007 EW information (Section \ref{sec: Prelim Sample}) and perform a linear regression where we treat the observed EW as a dependent variable and consider up to four SDSS colors as independent variables. We find that three colors ($g-r$, $r-i$, and $r-z$) are sufficient to produce a tight correlation ($u-r$ offers no additional benefits) against the logarithm of the EW:

\begin{equation}
\begin{split}
\log\, \mathrm {EW([OIII])} = 0.524\,(g-r)-0.210\,(r-i)- \\ 0.316\,(r-z)+2.164.
\end{split}
\label{eq:ew}
\end{equation}

\noindent The RMS of this relation is only 0.11 dex (Figure \ref{fig: ewoiii vs pred}). Note that the formula is by construction valid for $0.12<z<0.36$ GPs passing the C09 color selections, although it might work for a broader range of colors. Interestingly, adding the redshift to the regression tightens it only negligibly.


\subsection{Characteristics of Green Pea Candidates} \label{sec:char}

The top two panels in \autoref{fig: GPcans characteristics} show distributions of the GP candidates' $r$ magnitude (using linear and log scales, respectively) as compared to the {\it spectroscopic} GPs that were available in DR7 (and thus C09, orange histogram) and those currently present in DR18 (blue histogram). The gains in the numbers of GPs compared to spectroscopic samples are large and are not only limited to fainter GPs---we double the number of GPs compared to the DR18 spectroscopic GPs already at $r\sim 18.7$. We note tangentially that the original DR7 GPs do not go as faint as the ones that can now be spectroscopically selected in DR18. These differences between DR18 and DR7 likely stem from the changes in how the SDSS serendipitous spectroscopic targeting was carried out.

The bottom left panel of \autoref{fig: GPcans characteristics} shows the normalized distribution of the angular half-light radius ($R_{50}$) of our candidates compared to spectroscopic GPs. Our candidates have a broader range of sizes, extending to 1.35 arcsec, whereas spectroscopic ones do not go beyond 1.1 arcsec. We believe that these somewhat extended Peas are less prevalent in the spectroscopic sample because spectroscopic targeting for objects with GP-like colors often required the object to appear unresolved. We confirm this hypothesis by finding that 72\% of GP candidates are classified as non-stellar (\texttt{type = 3}) in SDSS, whereas this percentage is only 7\% in GPs with DR18 spectra. Our GP candidates seem to have uncovered a large population of GPs which, while still compact, are not exactly stellar-like, even in SDSS $\sim 1.3$-arcsec seeing imaging. For the GPs that arise from mergers, the ``Extended Peas" may represent a pre-coalescence stage. Indeed, we visually inspect the images of 41 GP candidates with $R_{50}>1.1$ arcsec that have been imaged in Subaru HSC DR3 (0.7 arcsec seeing, \citealt{Aihara22}), and find 3/4 to be close pairs or mergers. Some examples of the ``Extended Peas" are shown in Figure \ref{fig:Extended GPs}.   

We apply Eq.\ \ref{eq:ew} to estimate the [OIII]5007 EWs of newly identified GP candidates and compare it to known, spectroscopic GPs (bottom right panel of \autoref{fig: GPcans characteristics}). Again, we see that we have gains even among the most extreme systems. Out of 9628 GP candidates identified using the SED matching, 1501 have a predicted [OIII] EW above 500 \AA, which is some 5 times as many as the number of such sources in the DR18 spectroscopic GP sample. Gains become much larger at lower, but still relatively high EWs---a factor of 10 increase at [OIII] EW $\sim$ 300 \AA, the EW where the distribution peaks. Furthermore, unlike the EW distribution of spectroscopic GPs, the distribution of EWs of the GP candidates resembles a power law past the peak, i.e., for log EW[OIII] $\gtrsim$  2.5. This might hint at the underlying EW distribution of GPs also being a power law.

\subsection{Surface Density of Green Peas} \label{sec:density}

Here, we carry out the determination of the surface density of GPs for (a) the DR7 spectroscopically confirmed GPs that would have been available to C09, (b) the current (DR18) spectroscopically confirmed GPs, and (c) for the the newly identified GP candidates. Details of these samples are summarized in Table \ref{tab:samples}.

To determine the surface densities for (a) and (b), we cannot simply divide the number of objects by the official estimates for spectroscopic coverage areas. This is because the coverage that yields GPs may not encompass the entirety of the spectroscopic area. Indeed, inspecting the maps of spectroscopically identified GPs indicates that there are holes  within the areas nominally included in the spectroscopic coverage (e.g., in the Northern Galactic Cap, see the right panel in Figure \ref{fig: Positional cut galactic}). Therefore, we base the estimates on a relatively large region ($120 \degr <\alpha<160 \degr$, $20 \degr < \delta < 40 \degr$) which appears to be free of gaps and holes in terms of spectroscopically identified GPs. This region encompasses 689 sq.\ deg. The resulting surface densities are listed in Table \ref{tab:GPspatials}. We also estimate the effective areas of the sky that are surveyed for samples (a), (b) and (c) by multiplying the area of the gap-free region by the ratio of the total number of objects in a given sample to that in the gap-free region.

\begin{deluxetable}{ccc}
\tablecaption{Surface Densities of Green Peas}
\label{tab:GPspatials}     
\tablehead{
\colhead{Sample} & 
\colhead{Survey Area} &
\colhead{GP Surface Density} \\
\colhead{} &
\colhead{(sq.\ deg.)} &
\colhead{(per sq.\ deg.)}
}
\startdata                                      
DR7 spec (C09) & 4900 & 0.04 \\
DR18 spec & 6000 & 0.15 \\
DR18 phot & 10100 & 0.95 \\
\enddata
\end{deluxetable}
\vspace{-1\baselineskip}

We see that our new, photometrically selected sample yields 0.95 GP per sq.\ deg., which is some 25 times higher than the surface density of spectroscopic GPs in DR7, and 6 times higher than the surface density of spectroscopic GPs in DR18. 

Taking into account that the completeness of the color selection is about 50\% (see Section \ref{sec:disc}), the true density of $0.12<z<0.36$ compact EELGs in SDSS might be $\sim$ 2 GP per sq.\ deg.

\subsection{External Validation of Green Pea Candidate Selection}

In this section we are aiming to corroborate our estimates of high efficiency in identifying the GPs and low contamination rates using external (non-SDSS) information. 

\subsubsection{LAMOST Spectroscopic Survey}\label{sec:lamost}

LAMOST \citep{Wang96, SuCui04} has been performing a large-scale spectroscopic survey of the Northern sky utilizing a telescope with a 4-m effective aperture, located in Hebei, China. One of its programs was aimed at finding new GPs, following C09 color selection \citep{Liu22}. In addition to these dedicated searches, LAMOST was netting additional compact ELGs from other programs, similarly to how the GPs were included in SDSS in the first place. A sample of confirmed non-SDSS GPs from LAMOST would allow us to characterize the effectiveness of our SED matching classification method.

We match our pool of 18,880 photometric candidates to the LAMOST DR10 v2.0 database of low-resolution spectra using the online tool provided by the project. Matching is performed in 2 arcsec radii, and yields 771 spectra with redshifts belonging to 691 unique objects. The latest observing date is 2022 May 3. Out of 691 LAMOST matches, 117 already have SDSS spectra and redshifts, so we focus on 574 that do not.

The batch of 574 LAMOST objects includes 91 that the project has classified as stars, based on their spectra. Our photometric classification using the SED matching method agrees very closely, placing 84 of them in the stellar category. Next, there are 34 LAMOST objects classified as QSOs. Of them, our photometric classification places 23 into one of two QSO categories.

Finally, and most importantly, we look at LAMOST objects with galaxy spectra and redshifts in the GP range ($0.12<z<0.36$). There are 384 such objects without SDSS spectra. We classify (using the SED matching) 343 of them as Green Peas. The rest we mostly assigned star or QSO categories. Therefore, we conclude that the effectiveness with which SED matching identifies true GPs is $89\pm5\%$ (the error comes from the counting statistics). This is slightly higher, but within the Poisson error of the estimate based on self-validation ($84\pm 3\%$, Section \ref{sec: Refined sample}).

\begin{figure*}[htbp]
    \centering
    \includegraphics[width=0.9\textwidth,clip=true,trim={0 0 0 0}]{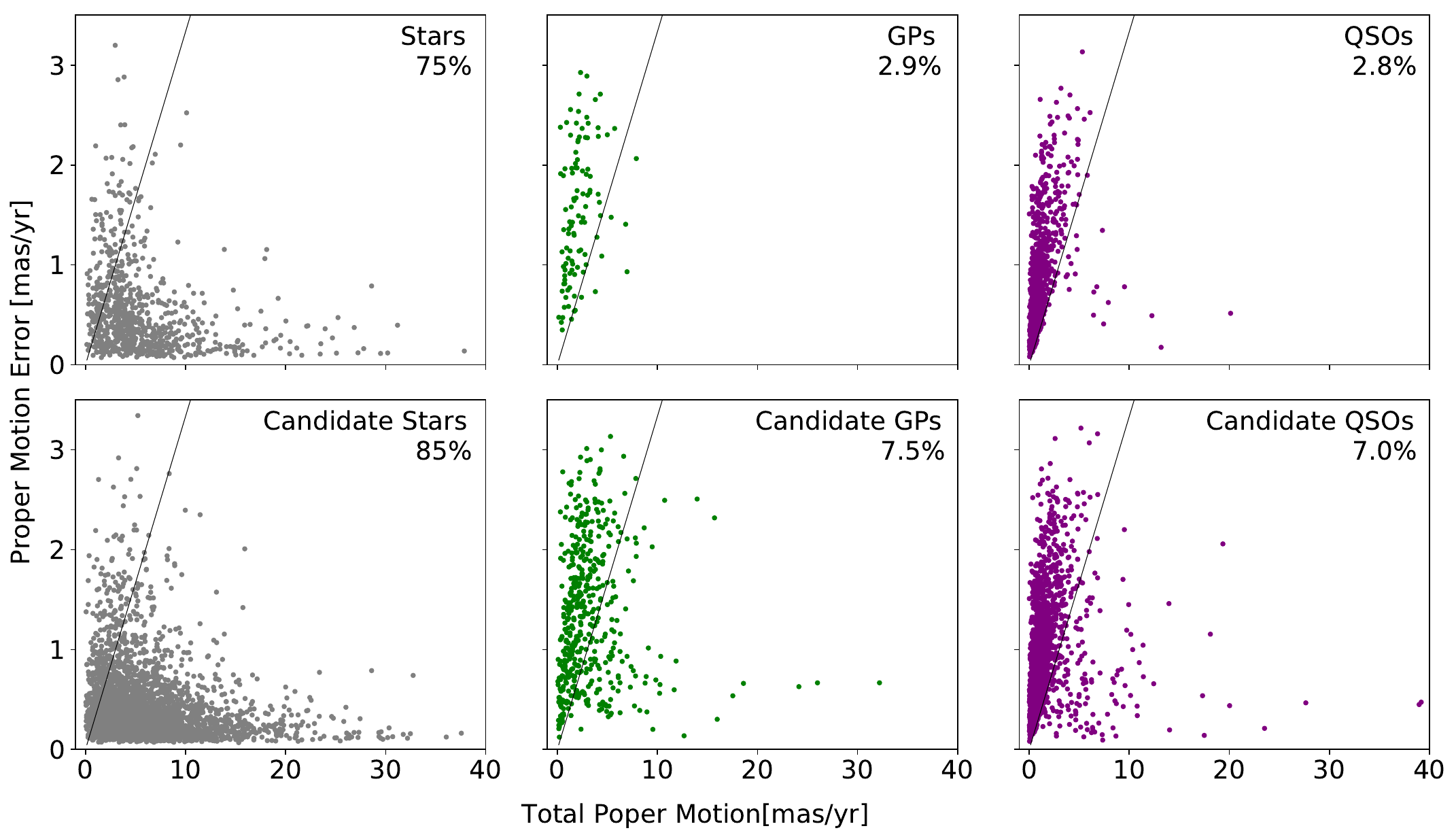}
    \caption{\textit{Gaia} DR3 proper motions for: stars, GPs and QSOs. Upper panels show spectroscopically confirmed classes, whereas the lower ones show the objects from the candidate pool that we classify using the SED matching method as stars, GPs and QSOs. Objects to the right of the line have significant ($> 3\sigma$) proper motions.}
    \label{fig: totpm vs pmerr}
\end{figure*}

\subsubsection{Green Peas from KISS Objective Prism Survey}

\citet{Brunker20} present 13 GPs selected from the KISS objective prism survey of emission line galaxies \citep{Salzer00}. In particular, their sample consists of spectroscopically confirmed [OIII]5007 emitters that fall in the SF region of the BPT diagram \citep{Salzer09}. Their GPs lie at $0.29<z<0.41$, so partially overlapping with the SDSS redshift window. KISS covers 128 sq.\ deg.\ and lies within the SDSS footprint.

We find that 4 of 13 KISS GPs are included in our list of 18,880 photometric candidates. Of the remaining nine, five lie at or above the redshift limit, one does not have a $W2$ detection and the remaining three do not pass C09 color cuts, although by a small margin. Of the four that we consider for our SED matching classification, three (KISSR 1508, 1516, 2005) are classified as GPs and one (KISSR 847) as a $z<2$ QSO. There is no indication in the spectrum presented by \citet{Brunker22} that this object has broad H$\beta$ expected in a QSO, but is rather a genuine GP, albeit one with exceptionally high [OIII]5007 EW of 1900 \AA.

While the sample is too small to draw any far-reaching conclusions, this exercise shows that our selection can correctly identify independently discovered GPs.

\subsubsection{\texorpdfstring{Green Peas from The H$\alpha$ Dot Survey}{Green Peas from The H-alpha Dot Survey}}
\label{sec:dots}

Our final external sample of GPs comes from \citet{Kimsey-Miller24}, which is based on the spectroscopically confirmed narrow-band selected compact emission-line sources from the H$\alpha$ Dot Survey \citep{Kellar12,Salzer20}.  Of 38 [OIII] emitters at $0.32<z<0.35$ that have SDSS photometry, \citet{Kimsey-Miller24} select 22 as GPs, using a $r-i<-0.1$ criterion. Their GPs tend to be fainter than our candidates, so only 10 pass our requirements for small photometric error and a WISE detection. Of them, five pass all the color cuts. Our SED matching classifies four as GPs (dots numbered 37, 119, 135 and 156) and one (dot 237) as a star. Again, this is a small sample, but points towards relatively high efficiency of our method.

\subsubsection{\textit{Gaia} Proper Motions}\label{sec:gaia_pm}

To further assess the reliability of our methodology, we utilize \textit{Gaia} DR3 astrometry \citep{GaiaCollab23} for proper motion analysis. Of 18,880 objects in the candidate pool, 7486 are cross-matched to \textit{Gaia} catalog, of which 5277 have proper motion information. \textit{Gaia} is complete to $r\sim 20$, and does not contain resolved sources, so it will contain only the brighter portion of our candidate pool. Of the 3696 objects within the candidate pool with spectroscopically known classes (Section \ref{sec: Prelim Sample}), 1426 have listed proper motions (114 GPs, 513 stars, and 799 QSOs). \autoref{fig: totpm vs pmerr} features the proper motion error versus the magnitude of the proper motion, separated by star, GP, and QSO (both low and high redshifts) for objects with spectroscopically known classification (upper panels) and for the candidates that we classified with the SED matching method (lower panels). Because the simple error propagation is not appropriate for components of a vector \citep{Lindegren22}, we take the  approximation of the total proper motion error to be the larger of the proper motion error in right ascension or declination. The black line represents the 3$\sigma$ limit---an object has significant proper motion if to the right of this line.

We see that, as expected, a much higher fraction of stars (75\%) show appreciable proper motion compared to the GPs or QSOs ($\sim 3$\% each). The similarity between the known stars' plot (top left panel in \autoref{fig: totpm vs pmerr}) and the candidate stars' plot (bottom left panel in \autoref{fig: totpm vs pmerr}), where even 85\% have appreciable proper motion, further reinforces the conclusion that our method by and large correctly classifies objects. Interestingly, several confirmed QSOs have proper motions in excess of $5\sigma$, and we could not identify the reasons behind it. The higher percentage of GPs with an appreciable proper motion among the GP candidates ($\sim 8$\%) than among the spectroscopically confirmed GPs ($\sim 3$\%) implies a stellar contamination rate of 5\%, in line with 4\% estimated based on the adjusted self-validation analysis ( \ref{sec:final}).

\section{Discussion} \label{sec:disc}

The principal goal of this paper is to significantly increase the number of identified GPs in the local universe over what is currently known, and thus provide a more complete census of these intriguing galaxies. In this section we aim to determine how far we have come towards achieving this goal.

In previous sections we have established that the effectiveness of our SED matching method is $\sim 85$\%. Effectiveness tells us what fraction of actual GPs is correctly identified as such, i.e., the recovery rate. This is different from an absolute assessment of the completeness, which would be the fraction of all GPs (whether they pass our candidate selection criteria or not) that we can identify as GPs. Any assessment of the completeness would obviously be performed against the GPs lying within our redshift window ($0.12<z<0.36$) and sky coverage. Secondly, here we are interested in a completeness with respect to the GPs that are actually detectable by SDSS imaging. We know from deep surveys of EELGs (e.g., \citealt{Amorin15,Kimsey-Miller24}) that there exist low-mass GPs that are too faint for SDSS. To objectively determine our completenes,s we would need an unbiased and complete dataset of high-EW galaxies from at least 100 sq.\ deg.\ (to have a sample big enough for a meaningful statistical estimate). No such dataset exists yet. Instead, we do the next best thing, which is to estimate the completeness from high-EW galaxies already in SDSS spectroscopic sample. While the analysis that follows based on GPs with SDSS spectra is not without its caveats (e.g., the sampling of GP parameter space is unlikely to be uniform), it still offers useful insights.

\begin{figure}[htbp]
    \centering
    \includegraphics[width=\columnwidth,clip=true,trim={8 8 8 8}]{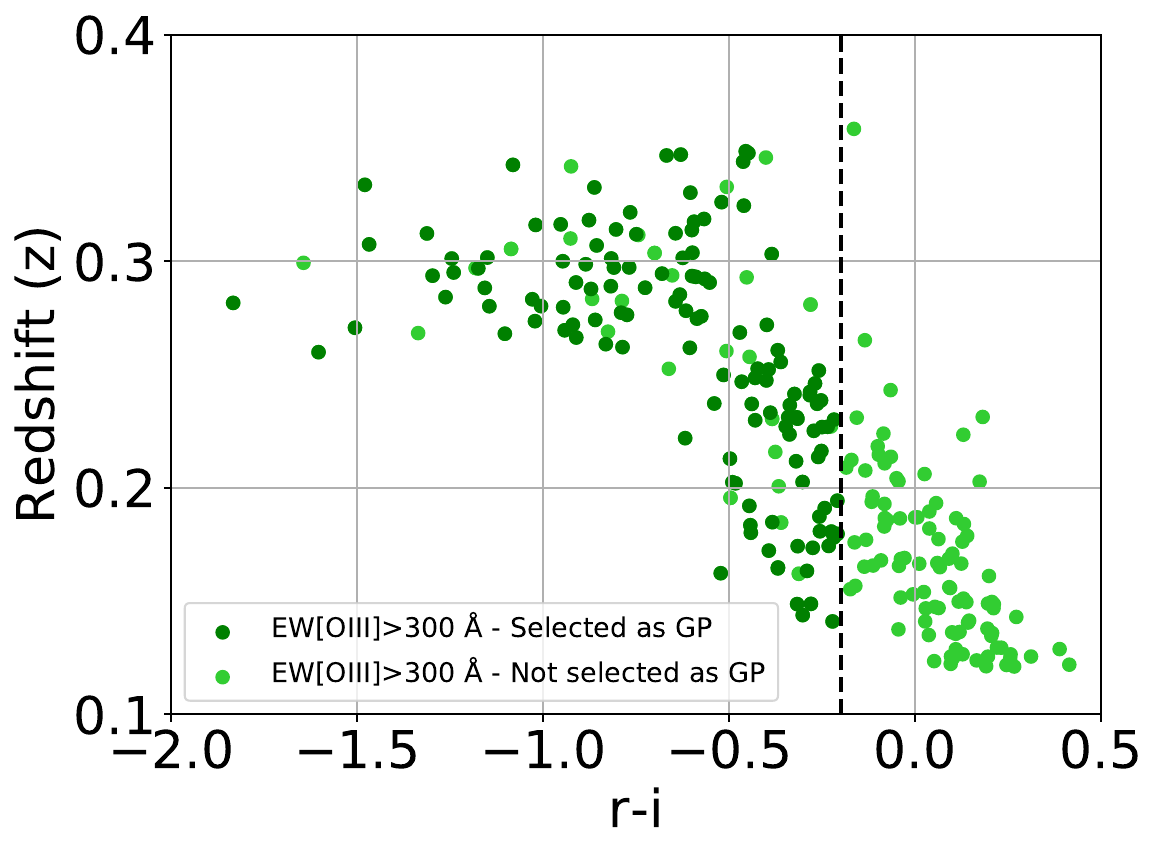}
    \caption{Assessment of the completeness. Redshift and $r-i$ color of all compact EELGs (EW[OIII]$>300$ \AA) which have EW measurements available in SDSS, overplotted with those that pass the selection cuts and are identified as Green Peas in this study. Higher incompleteness at $z<0.25$ is due to the H$\alpha$ line lying in the $i$ band and pushing the $r-i$ color towards the blue and thus outside of the selection cut.}
    \label{fig:z_ri}
\end{figure}

For the purposes of this exercise we will consider all galaxies with rest-frame EW[OIII]$>300$ \AA\ and $R_{50}<1.35$'' to be the target GP sample. What is important here is that no other criterion (besides the correct redshift range) is used. Namely, all other cuts discussed previously are ultimately the proxies for high [OIII]5007 EW or compact size, needed in the absence of spectra. It should also be pointed out that having an explicit cut on size might not even be necessary---objects with very high EW intrinsically tend to come only as very compact.

There are 256 target GPs as described above, of which 141 make it into the GP candidate pool, and 133 of which end up being classified by the SED matching method as GPs. This gives us an end-to-end completeness of 51\%.

Next, we ask what factors contribute to the completeness not being higher. We have found that the principal reason lies in part of the color selection of the GP candidate pool that involves $r-i$ color. This can be appreciated from Figure \ref{fig:z_ri}, where we show the redshift vs.\ the $r-i$ color of all 256 GPs, highlighting 133 that we identify as such (i.e., those that pass all selection cuts and are classified as GP by the SED matching method). The $r-i=-0.2$ cut, one of the color cuts used to select the candidates, is indicated. We see that above $z=0.25$ essentially all GPs are to the left of the cut and are identified as GPs in our analysis. Indeed, the end-to-end completeness of our GP identification at $z>0.25$ is a rather high (79\%). The remaining incompleteness at $z>0.25$ arises mainly because of the requirement to have WISE W1 and W2 photometry for the SED matching method.

Why is the completeness much lower at $z<0.25$ (35\%), or rather, why do many GPs below that redshift have colors redder than $r-i=-0.2$? The reason is that at these redshifts the H$\alpha$ line is still within the $i$-band, and being fairly strong, boosts the $i$ magnitude and makes $r-i$ color redder. An obvious remedy to this issue would be to drop in the future all selection cuts that involve $r-i$, namely:

\[
r - i \leq -0.2,
\]
\[
g - r \geq r - i + 0.5,
\]
\[
(r-i)_{\mathrm{fiber}} < 0
\]

\noindent Such modification would have kept 91\% of the parent sample (target GPs) in the candidate pool. However, the downside is that abandoning $r-i$-based cuts brings in a substantial additional contamination to the candidate pool from quasars, stars and low-EW compact galaxies---the very reason the cuts were introduced in the first place. The severity of that contamination might be such that even the SED matching method would not be able to overcome it, but we have not yet tested it.

To conclude, even with the 51\% overall completeness, we have vastly increased the number of identifiable GPs compared to what spectroscopic survey was providing. The completeness is much higher at $0.25<z<0.36$ (79\%). Future efforts will be needed to address the lower completeness at $0.12<z<0.25$.

\section{Green Pea Catalogs} \label{sec:cat}

To facilitate further studies, we present three lists of sources: 

\begin{enumerate}
  \setlength\itemsep{0em}
\item GPs confirmed with SDSS DR18 spectroscopy (917, Table \ref{tab:DR18 GPs})
\item GPs confirmed with non-SDSS spectroscopy (521, Table \ref{tab:nonSDSS GPs})
\item GP candidates (8313, Table \ref{tab:GPcans})
\end{enumerate}

For each source we list an ID, SDSS ObjID, position, $r$-band \texttt{modelMag} magnitude (de-reddened for the Milky Way dust extinction), Petrosian half-light radius based on $r$ band, predicted [OIII]5007 EW and, for lists (1) and (2), the redshift and its source. If additional SDSS data are required (e.g., full photometry) one can upload any of these three tables to SciServer CasJobs and run the SQL query given in Appendix \ref{app2}. We keep the columns of Tables \ref{tab:DR18 GPs} and  \ref{tab:nonSDSS GPs} identical to allow them to be easily concatenated into one table containing all spectroscopically confirmed GPs.

\textbf{GPs confirmed with SDSS DR18 spectroscopy.} Here we list 917 sources selected as GPs based on SDSS DR18 spectra. To recapitulate, the selection is based on C09 criteria (Eq.\ \ref{eq:c09} color cuts, a $0.12<z<0.36$ redshift window, and the requirement of a spectrum of galaxy type), to which we added photometric and size cuts (Eq.\ \ref{eq:crit}) that help eliminate low-EW interlopers. We do not require a WISE detection. This number is some 3.5 times higher than the original C09 GP sample, since it was based on DR7. We found that the increase of objects that would be selected as GPs was more or less gradual from one data release to the next, with the biggest increase being observed in DR16 (300 new GPs). The increase is likely due to the continued efforts to survey QSOs and target serendipitous objects, and not specifically from targeting GPs. For example, DR7 cataloged $\sim100,000$ quasars \citep{Schneider10}, while DR16 catalogues $\sim750,000$ \citep{Lyke20}.

\begin{deluxetable*}{lcccccccc}
\tablecaption{Green Peas confirmed with SDSS DR18 spectroscopy}
\label{tab:DR18 GPs}
\tablehead{
\colhead{ID} &
\colhead{SDSS ObjID} & 
\colhead{RA} & 
\colhead{DEC} & 
\colhead{$r$} & 
\colhead{$R_{50}$} & 
\colhead{log EW([OIII])} & 
\colhead{Redshift} & 
\colhead{Source} \\
\colhead{} & 
\colhead{(deg)} & 
\colhead{(deg)} & 
\colhead{(mag)} & 
\colhead{(arcsec)} & 
\colhead{(\AA)} & 
\colhead{} & 
\colhead{}
}
\startdata
GP-S0001 & 1237678579291914713 & 0.1336475 & 24.2718502 & 20.70 & 0.76 & 2.44 & 0.2672 & SDSS \\
GP-S0002 & 1237678661961056913 & 0.4584968 & 5.0125960  & 21.03 & 1.22 & 2.68 & 0.2868 & SDSS \\
GP-S0003 & 1237669680114499836 & 0.8672408 & 6.9391586  & 20.02 & 0.67 & 2.73 & 0.2757 & SDSS \\
GP-S0004 & 1237663307991548308 & 1.0928205 & 29.9217097 & 20.25 & 0.75 & 2.49 & 0.3056 & SDSS \\
GP-S0005 & 1237652900211720363 & 1.1263783 & -10.1915583 & 19.73 & 0.68 & 2.67 & 0.2386 & SDSS \\
\enddata
\tablecomments{Sample to show table content.  Coordinates, magnitude and half-light radius come from SDSS DR18. EW[OIII]5007 is an estimate based on Eq.\ \ref{eq:ew}. Source of redshift (in this case only SDSS) is indicated in the last column. GP-S in ID stands for Green Pea SDSS.}
\end{deluxetable*}

\textbf{GPs confirmed with non-SDSS spectroscopy.} This list of 521 sources is a subset of our photometrically-selected candidate pool (Sec.\ \ref{sec:final}, Table \ref{tab:samples}) for which non-SDSS spectroscopy confirming them as GPs exists. We identify these sources from the entire candidate pool (18,880 sources), and not just the sources that we classify as GP candidates using the SED matching (in the case some was misclassified as other than GP), although the large majority of them (88\%) was classified correctly.

Most of the GPs in list (2) come from LAMOST DR10 spectroscopic survey (384 objects, Sec.\ \ref{sec:lamost}). We require LAMOST redshift to be within the same window used for SDSS GPs ($0.12<z<0.36$) and a spectrum to be classified as a galaxy. LAMOST objects with SDSS spectra (107) are not included, since they are already in list (1).

Our next source of spectroscopic information is SIMBAD, which yields 84 sources not in list (1) and not in LAMOST. We select as GPs the SIMBAD sources with redshifts within the window, and with the primary classification being `G' (galaxy) or `EmG' (emission galaxy). We reject several sources with only photometric redshifts, and those classified as QSOs (5) or AGN (18). Although some of the AGN might be of type 2 AGN, and therefore acceptable as GPs, information regading their AGN type is not readily available. SIMBAD is of course a very heterogeneous compendium, but many of the 84 SIMBAD GP redshifts originally came from the 2QZ \citep{Croom04}, and the related 2SLAQ spectroscopic surveys of quasars \citep{Cannon06,Richards05}. The list includes 5 GPs from the H$\alpha$ Dot Survey (Sec.\ \ref{sec:dots}).

\textit{Gaia} has recently assembled an all-sky catalog containing $\sim 5$ million objects that might be galaxies based on the astrometric, photometric and BP/RP spectroscopic data and has provided redshift estimates for them \citep{Bailer-Jones23}. These galaxy candidates do not have to be extended, which means that the more compact GPs are not excluded a priori. Of the sources not already included based on SDSS, LAMOST or SIMBAD redshifts, we find 51 sources in the \textit{Gaia} Galaxy Candidates Catalog that have the Discrete Source Classifier classification \texttt{galaxy}. Given that \textit{Gaia} redshifts have a  relatively large $1\sigma$ error of 0.04, we were ready to accept sources within $2\sigma$ of our nominal redshift window, however only four objects exceeded it---one an obvious low-$z$ interloper, and 3 sources with \textit{Gaia} redshifts of up to 0.41. 

Our final additions to list (2) are two GPs from \citep{Brunker20}.

\begin{deluxetable*}{lcccccccc}
\tablecaption{Green Peas confirmed with non-SDSS spectroscopy}
\label{tab:nonSDSS GPs}
\tablehead{
\colhead{ID} &
\colhead{SDSS ObjID} & 
\colhead{RA} & 
\colhead{DEC} & 
\colhead{$r$} & 
\colhead{$R_{50}$} & 
\colhead{log EW([OIII])} & 
\colhead{Redshift} & 
\colhead{Source} \\
\colhead{} & 
\colhead{(deg)} & 
\colhead{(deg)} & 
\colhead{(mag)} & 
\colhead{(arcsec)} & 
\colhead{(\AA)} & 
\colhead{} & 
\colhead{}
}
\startdata
GP-N0001 & 1237679457610236177 & 1.8829384 & 18.4374339 & 19.97 & 0.92 & 2.62 & 0.2548 & LAMOST \\
GP-N0002 & 1237678778479083671 & 2.2685815 & 5.7600265  & 19.86 & 0.98 & 2.78 & 0.1700 & LAMOST \\
GP-N0003 & 1237679476398358848 & 4.4614953 & 21.6109234 & 20.33 & 0.72 & 2.71 & 0.2594 & LAMOST \\
GP-N0004 & 1237656496189342012 & 4.6285868 & 15.5160218 & 20.21 & 0.60 & 2.57 & 0.2912 & LAMOST \\
GP-N0005 & 1237678580904362300 & 4.7204556 & 25.6150833 & 19.85 & 1.09 & 2.57 & 0.2977 & LAMOST \\
\enddata
\tablecomments{Sample to show table content. Coordinates, magnitude and half-light radius come from SDSS DR18. EW[OIII]5007 is an estimate based on Eq.\ \ref{eq:ew}. Source of redshift is indicated in the last column. GP-N in ID stands for Green Pea Non-SDSS.}
\end{deluxetable*}

\textbf{GP candidates.} Our largest list by far (8313 objects), and the principal contribution of this study, consists of GP candidates without spectroscopic confirmation. The starting point for this list were 9628 objects from the candidate pool that we classified as GPs using SED matching (Sec.\ \ref{sec:final}, Table \ref{tab:samples}). From them, we obviously exclude 1102 spectroscopically confirmed GPs contained in lists (1) and (2). Additionally, spectroscopy that we discussed above allows us to eliminate 143 sources that are not GPs  (quasars, stars, and other objects outside of the redshift window). In practical terms, list (3) is constructed by removing all sources with redshifts from either SDSS, LAMOST, SIMBAD or \textit{Gaia}.

Furthermore, we remove 63 objects for which the \textit{Gaia} proper motions are significant at $>5\sigma$ level, i.e., $r<20$ star contaminants (Sec.\ \ref{sec:gaia_pm}). Finally, we visually inspect SDSS images of $\sim 70$ of the most extreme sources in terms of brightness, predicted EW and size, and remove 6 with spurious detections, bright backgrounds or imaging defects. The fraction of such sources is much lower overall. For example, we found only one in a random sample of 200.

\begin{deluxetable*}{lcccccc}
\tablecaption{Green Pea candidates (no spectroscopic confirmation)}
\label{tab:GPcans}
\tablehead{
\colhead{ID} &
\colhead{SDSS ObjID} & 
\colhead{RA} & 
\colhead{DEC} & 
\colhead{$r$} & 
\colhead{$R_{50}$} & 
\colhead{log EW([OIII])} \\
\colhead{} & 
\colhead{(deg)} & 
\colhead{(deg)} & 
\colhead{(mag)} & 
\colhead{(arcsec)} & 
\colhead{}
}
\startdata
GP-C0001 & 1237678598090457376 & 0.0190789 & 3.1667026  & 20.05 & 0.77 & 2.47 \\
GP-C0002 & 1237680332713952145 & 0.1195483 & 29.9815122 & 21.24 & 0.75 & 2.46 \\
GP-C0003 & 1237663278464696570 & 0.1838829 & 0.8578103  & 20.30 & 0.61 & 2.52 \\
GP-C0004 & 1237679323919876841 & 0.2180723 & -4.4281471 & 20.78 & 0.68 & 2.51 \\
GP-C0005 & 1237663277927825747 & 0.2208775 & 0.4583196  & 21.01 & 0.78 & 2.39 \\
\enddata
\tablecomments{Sample to show table content. Coordinates, magnitude and half-light radius come from SDSS DR18. EW[OIII]5007 is an estimate based on Eq.\ \ref{eq:ew}. GP-C in ID stands for Green Pea Candidate.}
\end{deluxetable*}

\section{Summary}

In this work, we present a new, highly efficient photometric method that hugely increases the number of extreme emission line galaxies (``Green Peas") identifiable from SDSS compared to what can be selected based on spectra. The method is designed to work in the same volume that was probed by C09 ($0.12<z<0.36$) and utilizes their color selection as the starting point. Our main findings are:

\begin{enumerate}
  \setlength\itemsep{0em}
    \item Spectroscopically identified GPs in SDSS are a minor portion of the overall population of GPs, even at bright magnitudes.
    \item In the absence of spectroscopy, the selection of $0.12<z<0.36$ GPs using only the C09 color criteria (Eq.\ \ref{eq:c09}) is ineffective because these photometric candidates are dominated by contaminants (stars, quasars and HII regions of nearby ($z\lesssim0.03$) galaxies).
    \item Introduction of new photometric and compactness cuts (Eq.\ \ref{eq:crit}) can remove $\sim 70$\% of contamination from color-selected photometric candidates.
    \item Our new SED matching technique, in which we compare the optical magnitude and optical and near-IR colors of photometric candidates to spectroscopic objects of known nature (GP, star, quasar, non-GP galaxy or HII region) is able to remove most of the remaining contaminants and identify likely GPs. Inclusion of near-IR bands from WISE (W1 and W2) improves the ability to correctly identify GPs.
    \item The SED matching method recovers 84\% of known (spectroscopic) GPs, and has a contamination rate (objects not really GPs) of $\sim 10$\%. 
    \item The method identifies $\sim 9600$ GP candidates in SDSS DR18---10 times more than the number of spectroscopically confirmed GPs, and 44 times more than the number of spectroscopically selected GPs available originally. Of these, $\sim 8300$ are new, without spectra from SDSS, LAMOST or other sources.
    \item New GP candidates span a full range of magnitudes, including many as bright as $r \sim 19$. Of them, 1200 are predicted to have very high [OIII]5007 equivalent widths ($>500$ \AA).
    \item The sizes of GP candidates extend beyond the ones with SDSS spectra. A high fraction of  such ``Extended Peas" ($R_{50}>1.1$ arcsec), which look like one source in SDSS, are resolved as multiple sources in Subaru imaging, suggesting that these GPs are mergers in pre-coalescence stage. 
    \item The surface density of $0.12<z<0.36$ GPs in SDSS is $\sim$2 per sq.\ deg., with the completeness of the selection method taken into account. . 
    \item The [OIII]5007 equivalent widths for $0.12<z<0.36$ GPs can be predicted from a combination of SDSS broad-band colors, with an error of only 0.11 dex.
\end{enumerate}

Finally, we provide listings of 917 $0.12<z<0.36$ GPs confirmed using current (DR18) SDSS spectroscopy, 521 GPs with spectroscopic redshifts from LAMOST and other sources, and 8313 newly identified GP candidates that await spectroscopic confirmation.

The new large sample of GP candidates presented here paves the way for a range of studies focusing on these intriguing galaxies.


\section*{Acknowledgments}

Heather Samonski gratefully acknowledges the support from the John and A-Lan Reynolds Post-Baccalaureate Fellowship. The authors thank the reviewers for making helpful suggestions for improving the paper.

Funding for the Sloan Digital Sky Survey V has been provided by the Alfred P. Sloan Foundation, the Heising-Simons Foundation, the National Science Foundation, and the Participating Institutions. SDSS acknowledges support and resources from the Center for High-Performance Computing at the University of Utah. SDSS telescopes are located at Apache Point Observatory, funded by the Astrophysical Research Consortium and operated by New Mexico State University, and at Las Campanas Observatory, operated by the Carnegie Institution for Science. The SDSS web site is \url{www.sdss.org}.

SDSS is managed by the Astrophysical Research Consortium for the Participating Institutions of the SDSS Collaboration, including Caltech, The Carnegie Institution for Science, Chilean National Time Allocation Committee (CNTAC) ratified researchers, The Flatiron Institute, the Gotham Participation Group, Harvard University, Heidelberg University, The Johns Hopkins University, L'Ecole polytechnique f\'{e}d\'{e}rale de Lausanne (EPFL), Leibniz-Institut f\"{u}r Astrophysik Potsdam (AIP), Max-Planck-Institut f\"{u}r Astronomie (MPIA Heidelberg), Max-Planck-Institut f\"{u}r Extraterrestrische Physik (MPE), Nanjing University, National Astronomical Observatories of China (NAOC), New Mexico State University, The Ohio State University, Pennsylvania State University, Smithsonian Astrophysical Observatory, Space Telescope Science Institute (STScI), the Stellar Astrophysics Participation Group, Universidad Nacional Aut\'{o}noma de M\'{e}xico, University of Arizona, University of Colorado Boulder, University of Illinois at Urbana-Champaign, University of Toronto, University of Utah, University of Virginia, Yale University, and Yunnan University.

This publication makes use of data products from the Wide-field Infrared Survey Explorer, which is a joint project of the University of California, Los Angeles, and the Jet Propulsion Laboratory/California Institute of Technology, funded by the National Aeronautics and Space Administration.

This research has made use of the SIMBAD database, CDS, Strasbourg Astronomical Observatory, France. 

Guoshoujing Telescope (the Large Sky Area Multi-Object Fiber Spectroscopic Telescope LAMOST) is a National Major Scientific Project built by the Chinese Academy of Sciences. Funding for the project has been provided by the National Development and Reform Commission. LAMOST is operated and managed by the National Astronomical Observatories, Chinese Academy of Sciences.

This work has made use of data from the European Space Agency (ESA) mission {\it Gaia} (\url{https://www.cosmos.esa.int/gaia}), processed by the {\it Gaia} Data Processing and Analysis Consortium (DPAC, \url{https://www.cosmos.esa.int/web/gaia/dpac/consortium}). Funding for the DPAC has been provided by national institutions, in particular the institutions participating in the {\it Gaia} Multilateral Agreement.


\appendix
\section{SQL Query for Selecting the Green Peas Candidate 
Pool}\label{app}

\begin{verbatim}
SELECT count(*) -- replace "count(*)" with specific fields
    
FROM PhotoPrimary as p 
  JOIN Run as r on p.Run = r.Run
  JOIN wiseForcedTarget as w on p.ObjID = w.ObjID

WHERE

--Valid magnitude range
dered_u>15 and dered_u<24 and 
dered_g>15 and dered_g<24 and 
dered_r>15 and dered_r<24 and 
dered_i>15 and dered_i<24 and 
dered_z>15 and dered_z<24 and 

--Reasonable mag errors
modelMagErr_u<0.3 and 
modelMagErr_g<0.3 and 
modelMagErr_r<0.3 and 
modelMagErr_i<0.3 and 
modelMagErr_z<0.3 and 

--C09 color selection
dered_u - dered_r <= 2.5 and 
dered_r - dered_i <= -0.2 and 
dered_r - dered_z <= 0.5 and 
dered_g - dered_r >= dered_r - dered_i + 0.5 and 
dered_u - dered_r >= 2.5*(dered_r - dered_z) and

--Four new cuts
fiberMag_r-extinction_r - (fiberMag_i-extinction_i) < 0 and
dered_r > 17.5 and
petroR50_r < 1.35 and
fiberMag_r-modelMag_r > -0.3 and fiberMag_r-modelMag_r < 0.7 and 

--unWISE detection
w1_mag > 10 and w1_mag < 30 and
w2_mag > 10 and w2_mag < 30 and

--Exclude Galactic plane and isolated SEGUE stripes
stripe <=86 and
(b >20 or b <-20)
\end{verbatim}

\section{SQL Query for Obtaining Additional SDSS Data for Green Pea Catalogs}\label{app2}

\begin{verbatim}
SELECT p.objid,p.dered_u,p.dered_g,p.dered_r,p.dered_i,p.dered_z
-- add or replace with fields from PhotoPrimary

FROM MyDB.uploaded_table as my
  LEFT JOIN PhotoPrimary AS p on p.ObjID = my.ObjID
\end{verbatim}

\bibliography{main}{}
\bibliographystyle{aasjournal}

\end{document}